\documentclass[aps,prd,preprint,showpacs]{revtex4}
\usepackage{graphics}% Include figure files
\usepackage{dcolumn}% Align table columns on decimal point
\usepackage{bm}% bold math

\begin{document}

\title{Preferred-Frame and $\bm{CP}$-Violation Tests with Polarized Electrons}
\author{B. R. Heckel}
\author{E. G. Adelberger}
\author{C. E. Cramer}\altaffiliation[Current address: ]{Harvard-Smithsonian Center for Astrophysics}
\author{T. S. Cook}
\author{S. Schlamminger}
\author{U. Schmidt}\altaffiliation[Current address: ]{Physicalisches Institute, Universit\"at Heidelberg%, Philosophenweg 12, D-69120 Germany
}

\affiliation{Center for Experimental Nuclear Physics and Astrophysics, Box 354290,
University of
Washington, Seattle, WA 98195-4290}
\date{\today}

\begin{abstract}
We used a torsion pendulum containing $\approx 10^{23}$ polarized electrons to 
search new interactions that couple to electron spin. We limit CP-violating interactions between the pendulum's electrons and unpolarized matter in the earth or the sun, 
test for rotation and boost-dependent preferred-frame effects using the earth's rotation and velocity with respect to the entire cosmos, 
and search for exotic velocity-dependent potentials between polarized electrons and unpolarized matter in the sun and moon. 
We find $CP$-violating parameters $|g_{\rm P}^e g_{\rm S}^N|/(\hbar c)< 9.4 \times 10^{-37}$ and $|g_{\rm A}^e g_{\rm V}^N|/(\hbar c) < 1.2\times 10^{-56}$ for $\lambda > 1$AU.  We test for preferred-frame interactions of the form $V= -\bm{\sigma}^e \!\cdot\! \bm{A}$, $V=-B\bm{\sigma}^e \!\cdot\!\bm{v}/c$, or $V=-\sum \sigma^e_i C_{ij} v_j/c$, where $\bm{v}$ is the velocity of the Earth with respect to the CMB restframe and $i,j$ represent the equatorial inertial coordinates $X$,$Y$ and $Z$. We constrain all 3 components of $A$, obtaining $1\sigma$ upper limits $|\bm{A}_{X,Y}| \leq 1.5 \times 10^{-22}$~eV and $|\bm{A}_{Z}| \leq 4.4 \times 10^{-21}$~eV that 
may be compared to the benchmark value $m_e^2/M_{\rm Planck}= 2 \times 10^{-17}$~eV. 
Interpreting our constraint on $\bm{A}$ in terms of non-commutative geometry, we obtain an upper bound of 
$(355~ l_{\rm GUT})^2$ 
on the minimum observable area, where $l_{\rm GUT}= \hbar c/(10^{16}$~GeV)
is the grand unification length. We find that $|B|\leq 1.2\times 10^{-19}$~eV. 
All 9 components of $\bm{C}$ are constrained at the $10^{-17}$ to $10^{-18}$ eV level. We 
%also give results %for $\bm{A}$ and $\bm{C}$ when the velocity is taken to be the earth's velocity around the sun; these 
determine 9 linear combinations of parameters of the Standard Model Extension; rotational-noninvariant and boost-noninvariant terms are limited at roughly the $10^{-31}$ GeV and $10^{-27}$ GeV levels, respectively.  Finally, we find that the gravitational mass of an electron spinning toward the galactic center differs by less than about 1 part in $10^{21}$ from an electron spinning in the opposite direction. As a byproduct of this work, the density of polarized electrons in
Sm$\,$Co$_5$ was measured to be $(4.19\pm 0.19)\times 10^{22}\;{\rm cm}^{-3}$ at a field of 9.6 kG.
\end{abstract}
\maketitle
\section{Introduction}
\label{sec: intro}
This paper describes constraints on possible new spin-coupled interactions using a torsion  
pendulum containing $\approx 1 \times 10^{23}$ polarized electrons. Several rather different considerations motivated this work. General relativity (the classical theory that forms the standard model of gravity) does not in itself make predictions about the gravitational properties of intrinsic spin--a quantum mechanical effect with no classical analog. Because the classic equivalence-principle and inverse-square-law experiments all used unpolarized test bodies they can shed no light on this issue. In addition, these ``fifth force'' experiments were completely insensitive to the purely spin-dependent forces arising from the first-order exchange of unnatural parity ($0^-$, $1^+$, etc) bosons. Lastly, the spin pendulum provided a means to search for a new class of preferred-frame effects that involve intrinsic spin. 

We established very
tight bounds on long-range $CP$-violating interactions and on velocity-dependent forces by searching for the effects of unnatural-parity bosons exchanged between the electrons in our pendulum and unpolarized
matter in the earth and in the sun. Similarly, we made sensitive searches for preferred-frame effects defined by the entire cosmos. In this case, we checked whether the spins in our pendulum preferred to orient themselves in a direction fixed in inertial space, or if they had a generalized helicity defined by their velocity with respect to the rest-frame of the cosmic microwave background.
Finally, we investigated non-commutative space-time geometries and obtained constraints that lie far beyond the reach of any proposed accelerator. In each case, our bounds are interesting because of their high sensitivity. 
A Letter on the first results from this apparatus has already appeared\cite{he:06}. This paper, which supercedes Ref.~\cite{he:06}, reports additional data and more powerful constraints, provides a complete account of the experimental work, and presents a more extensive discussion of our constraints. 
\subsection{Cosmic Preferred Frames}
\subsubsection{General Considerations}
It seems natural to us, when testing possible Lorentz-violating preferred-frame scenarios, to let the
cosmic microwave background (CMB) define the fundamental inertial frame.  We consider, in turn, three types of preferred-frame effects involving spin. First, we consider an interaction that violates rotational invariance
and produces a laboratory potential
\begin{equation}
V=-\bm{\sigma} \cdot {\bm A}~,
\label{eq: def A}
\end{equation}
where $\bm{A}$ is a preferred direction with respect to the cosmos as a whole 
Next, we consider 
%a specific case of Eq.~\ref{eq: def C}, 
a helicity-generating interaction that violates boost invariance as well, {\em i.e.} a lab-frame interaction 
\begin{equation}
V=-B \bm{\sigma} \cdot {\bm v}/c~,
\label{eq: def B}
\end{equation}
where $\bm{v}$ is the velocity of the spin with respect to the CMB rest-frame. This velocity has two significant components--the velocity of the earth with respect to the sun ($v_{\odot}/c\sim 10^{-4}$) and the velocity of the sun with respect to the CMB rest frame ($v_{\rm CMB}/c \sim 10^{-3}$); the small velocity due to the earth's rotation is neglected. 

Finally, following the general ideas of 
Kosteleck\'y and coworkers\cite{co:97}, we consider a tensor helicity-generating term that produces a laboratory interaction
\begin{equation}
V=-\sum_{i,j} \sigma_i \frac{v_j}{c} C_{ij}~.
\label{eq: def C}
\end{equation}
We take Cartesian equatorial coordinates for our inertial frame, where $\hat{\bm Z}$  points North along the earth's spin axis,
$\hat{\bm X}$ points from the earth to the sun at the 2000 vernal equinox, and $\hat{\bm Y}=\hat{\bm Z} \times \hat{\bm X}$.

These possible spin-dependent cosmic 
preferred-frame effects differ from the known CMB preferred-frame effect in two important ways. First, if $\bm{A}$ is a time-even polar vector, $B$ a time-even scalar and 
$\bm{C}$ a time-even tensor (untestable assumptions even in principle), then the interactions in 
Eqs.~\ref{eq: def A}, and either \ref{eq: def B} or \ref{eq: def C} violate time-reversal and parity, respectively. Second, if such spin-dependent preferred-frame effects exist, they could be observed within an opaque box that shielded out any normal information about the outside world.
Although we naively expect that any such preferred-frame effects
to be suppressed by the Planck scale and therefore of order $m_e^2/M_{\rm Planck}\approx 2 \times 10^{-17}$~eV, our results are sufficiently sensitive to probe such tiny effects.
\subsubsection{The Standard Model Extension effective theory}
Kosteleck\'y and coworkers\cite{co:97} have developed a preferred-frame scenario in which spin-1 and spin-2 fields were spontaneously generated in the early universe and were subsequently inflated to enormous extents. These fields clearly violate Lorentz symmetry but in a very gentle way. Rotations and boosts of an observer remain Lorentz invariant, but the same operations on a particle (but not the Universe) obviously do not respect the invariance principle.
The Lorentz non-invariance invalidates the Pauli-Luders theorem, allowing Kosteleck\'y et al.
to construct a field theory with $CPT$-violating
effects (the
Standard-Model Extension or SME) that has been widely used to quantify the
sensitivity of various $CPT$ and preferred-frame tests involving photons, leptons, mesons and baryons (Ref.~\cite{ko:08} has an extensive set of references to this work).

Our $\bm{A}$ coefficient is identical to the SME coefficient $\bm{\tilde{b}}^e$
which contains contributions from $CPT$-violating as well as $CP$-violating terms\cite{bl:00}.
The SME has similar terms to our $\bm{C}$ coefficients, but they are conventionally expressed in sun-fixed coordinates. To facilitate SME analyses and to compare our results with other work based on SME analyses, we also quote results in which the sun's velocity with respect to the CMB rest frame is ignored.
\subsubsection{The Ghost Condensate dynamical theory}
Arkani-Hamed and colleagues have developed a consistent dynamical mechanism for the spontaneous breaking of Lorentz symmetry that, in effect, proposes a pervasive, massless, aether-like fluid consisting of Nambu-Goldstone bosons associated with the broken time-diffeomorphism symmetry\cite{ar:04,ch:06}. The bosons form a ``ghost condensate''  because the assumption of translational invariance demands a negative kinetic term in the Lagrangian. They form a condensate because higher-order terms are needed to stabilize the negative kinetic term. This fluid has remarkable properties. Unlike a classical aether, it can be excited. It modifies Newtonian gravity at long ranges and late times in the evolution of the universe. It behaves like a cosmological constant and is capable of accelerating the expansion of the universe, but maximally violates the Equivalence Principle. It defines a preferred frame that converges with the rest frame of the CMB as it reaches equilibrium, and as it evolves to its final state it mimics dark matter. Cosmological tests of gravity limit the diffeomorphism symmetry-breaking scale $M$ to be less than 10 MeV. Interesting physics is possible within this constraint; if $M$ were $\sim\! 10^{-3}$~eV the condensate could drive the observed acceleration of the universe; if $M$ were $\sim\! 1$~eV it would mimic the observed dark matter. 
Particularly interesting in the present context, 
a fermion moving with respect to the ghost condensate experiences an interaction
\begin{equation}
V=\frac{M^2}{F} \, \bm{\sigma} \cdot \bm{v}/c~,
\label{eq: Thaler mu}
\end{equation}
where $F$ is a mass scale associated with the coupling of the fermion to the condensate, and $\bm{v}$ is the fermion's velocity with respect to the condensate, which we assume to be at rest in the frame where the CMB is essentially isotropic. Equation~\ref{eq: Thaler mu} has the same form as Eq.~\ref{eq: def B}.
\subsection{Non-commutative spacetime geometries}
Preferred-frame effects also occur in non-commutative space-time geometries\cite{ca:01,hi:02} that can arise in $D$-brane models. In these geometries, the space-time coordinates $x_{\mu}$
do not commute, but instead satisfy
\begin{equation}
\left[\hat{x}_{\mu},\hat{x}_{\nu}\right]=i\Theta_{\mu\nu}~,
\label{eq: non-comm def}
\end{equation}
where $\Theta_{\mu\nu}$ is a real, antisymmetric tensor that has dimensions of length squared and $|\Theta|$ represents the smallest ``observable'' patch of area. 
It is often assumed that $\Theta_{0i}=0$ to avoid problems with causality. We make the usual additional assumption that $\Theta_{\mu\nu}$ is constant over the space-time region spanned by our experiment, so that it defines a preferred direction $\eta^i= \epsilon^{ijk}\Theta_{jk}$. 
One consequence of Eq.~\ref{eq: non-comm def} is that a spinning electron experiences a $CP$-violating interaction\cite{hi:02,an:01} 
\begin{equation}
{\cal L}_{\rm eff}=\frac{3}{4} m \Lambda^2 \left( \frac{\alpha}{4 \pi \hbar} \right)^2 \Theta^{\mu\nu} \bar{\psi}\sigma_{\mu\nu} \psi~,
\label{eq: non-comm effect}
\end{equation}
where $\Lambda$ is a cutoff conventionally taken to be $\sim 1$ TeV, and $\alpha$ is the fine structure constant.
Equation~\ref{eq: non-comm effect} is equivalent to
\begin{equation}
V_e=\bm{\sigma}_e \cdot \bm{\hat{\eta}}\left(\frac{\Lambda}{1~{\rm TeV}}\right)^2  3.33\times 10^{36}~\frac{\rm eV}{\rm m^2}\:|\Theta|~ ,
\label{eq: non-com pot}
\end{equation}
which has the same general form as Eq.~\ref{eq: def A}.

\subsection{Forces from exotic boson exchange}
\label{sec: pseudo}
The exchange of pseudoscalar particles produces spin-dependent forces that vanish between unpolarized bodies. As a consequence, there are few experimental constraints on such forces. Moody and Wilczek\cite{mo:84} discussed the forces produced by the exchange of low-mass, spin-0 particles and
pointed out that particles containing $CP$-violating $J^{\pi}=0^+$ and $J^{\pi}=0^-$ admixtures would produce a macroscopic, $CP$-violating ``monopole-dipole'' interaction between a polarized electron and an unpolarized atom with mass and charge numbers $A$ and $Z$ 
\begin{equation}
V_{e A}(r)= g_{\rm P}^e g_{\rm S}^A  \frac{\hbar}{8 \pi m_e c}{\bm{\sigma}}_e \cdot \left[ {\bm{\hat r}} \left( \frac{1}{r \lambda}+\frac{1}{r^2}\right)e^{-r/\lambda}\right]~, 
\label{eq: MW}
\end{equation} 
where $m_{\phi}=\hbar/(\lambda c)$ is the mass of the hypothetical spin-0 particle, $g_{\rm P}$ and $ g_{\rm S}$ are its pseudoscalar and scalar couplings, and $g_{\rm S}^A= Z(g_{\rm S}^e+g_{\rm S}^p) + (A-Z) g_{\rm S}^n$. For simplicity, we assume below that $g_{\rm S}^p=g_{\rm S}^n=g_{\rm s}^N$ and $g_{\rm S}^e=0$ so that $g_{\rm S}^A=A g_{\rm S}^N$; constraints for other choices of the scalar couplings can be readily obtained by scaling our limits. 

Recently Dobrescu and Mocioiu\cite{do:06} classified the kinds of potentials, constrained only by rotational and translational invariance, that might arise from exchange of low-mass bosons. Our work is sensitive to 3 of their potentials; in addition to a potential equivalent to Eq.~\ref{eq: MW}, we probe the two velocity-dependent potentials    
\begin{equation}
V_{e N}(r)=\frac{\bm{\sigma}_e}{8 \pi} \cdot \left[f_{\perp}\frac{\hbar}{c}\frac{(\bm{\tilde{v} \times {\hat r}}
)}{m_e}\left( \frac{1}{r \lambda}+\frac{1}{r^2}\right) + 
f_v \frac{\bm{\tilde{v}}}{r}
\right]e^{-r/\lambda}~,
\label{eq: BD}
\end{equation}
where $\bm{\tilde {v}}$ is the relative velocity in units of $c$. Both terms can be generated by one-boson exchange in Lorentz-invariant theories.
The parity-conserving $f_{\perp}$ term can arise from
scalar or vector boson exchange, while the parity-violating $f_v$ term
can be induced by spin-1 bosons that
have both vector and axial couplings to electrons, giving $f_v=2g_{A}^e g_{V}^N$. We constrain the parameters of Eq.~\ref{eq: BD} by studying the interaction between the polarized electrons in our pendulum and unpolarized nucleons in the sun and moon.

\section{Apparatus}
\label{sec: apparatus}
\subsection{Rotating torsion balance}
\label{subsec: balance}
\begin{figure}[t]
\hfil\scalebox{.7}{\includegraphics*[201pt,35pt][430pt,336pt]{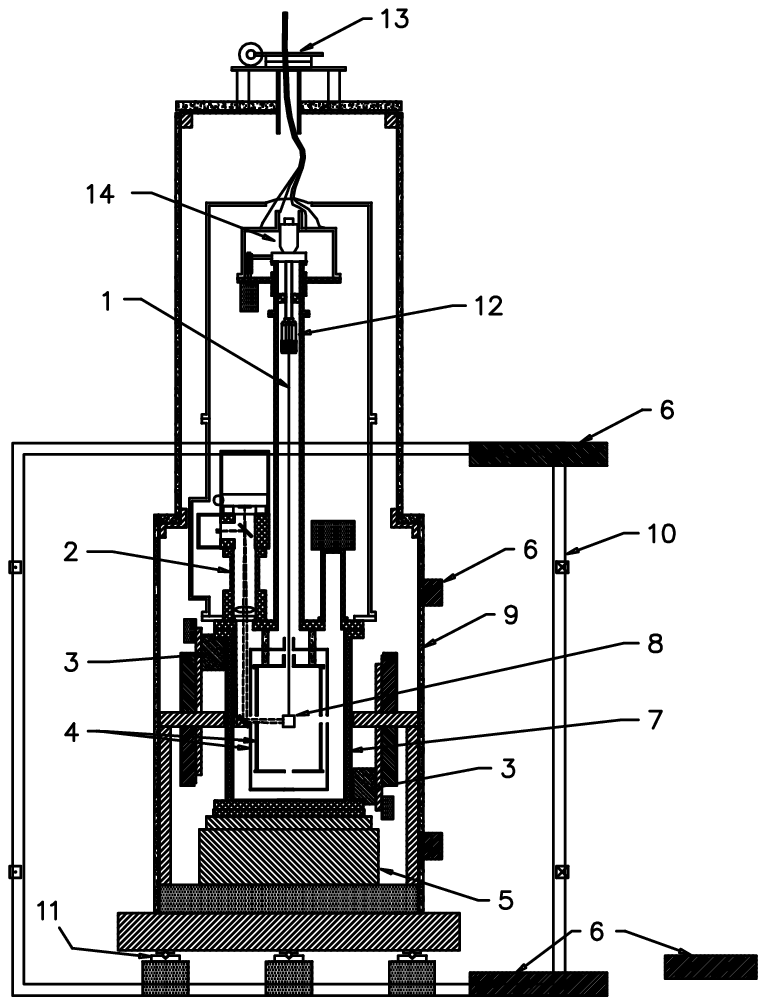}}\hfil
\caption{Scale cross-section of the E\"ot-Wash rotating torsion balance; 1: torsion fiber, 2: autocollimator, 3: $Q_{21}$
gravity-gradient compensator, 4: nested magnetic and electrostatic shields, 5: turntable, 6: $Q_{31}$ gravity-gradient compensator, 7: vacuum vessel, 8: spin pendulum, 9: thermal shield, 10: 3-axis Helmholtz coil system, 11: thermally controlled feet, 12: mode spoiler, 13: co-rotating slip-ring assembly, 14: upper fiber attachment mechanism. Two additional layers of magnetic shielding (not shown for clarity) are immediately inside and outside of the vacuum vessel.}
\label{fig: EW balance}
\end{figure}

This measurement used a substantially upgraded version of the E\"ot-Wash rotating torsion balance that was 
used for a previous test\cite{ba:99} of the Equivalence Principle. This device, in turn, was an improved version of the instrument described in detail in Ref.~\cite{su:94}. The essential features of this balance are shown in Fig.~\ref{fig: EW balance}. Briefly, a vacuum vessel containing a torsion pendulum and its associated optical readout system was rotated uniformly about a vertical axis. The pendulum hung in a region of very low magnetic fields and gravity gradients, and the entire instrument was
temperature-controlled (stability $\pm 5$ mK). 
Magnetic fields and gradients were reduced by a stationary set of Helmholtz coils and 4 layers of co-rotating mu-metal shielding. Gravity gradients were canceled as described in Ref.~\cite{su:94}, with a precision that was limited by the fluctuating water content of the ground outside our laboratory.
%
%	Figure 1
%

For this measurement, we made several substantial improvements to the instrument used in Ref.~\cite{ba:99}.
The turntable was upgraded by a ``feet-back'' system, described in more detail in Appendix~\ref{ap: feetback}, that kept its rotation axis vertical to better than 10 nanoradians. 
The autocollimator system that monitored the pendulum twist was upgraded from that described in Ref.~\cite{su:94} by using the ``two bounce'' geometry discussed in Ref.~\cite{ho:04}. We also improved the resolution of the temperature monitoring system and upgraded the co-rotating magnetic shielding.

\begin{figure}[t]
\hfil\scalebox{.46}{\includegraphics*[58pt,37pt][577pt,448pt]{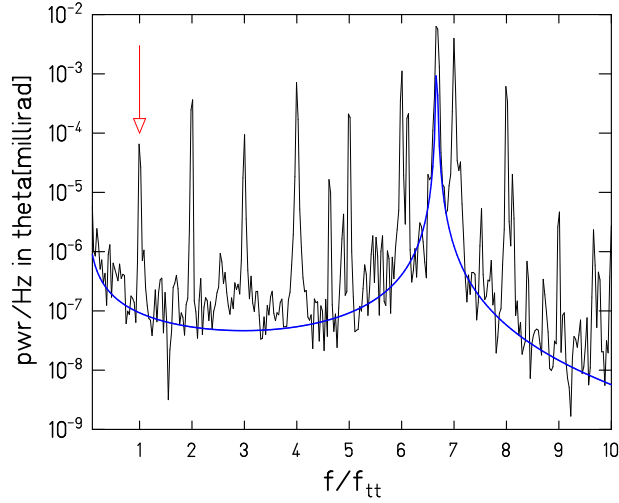}}\hfil
\caption{Spectral power density in the twist signal. The horizontal scale shows the frequency in units of the turntable rotation frequency. The smooth curve shows the expected thermal noise density for
this pendulum which had $Q\approx 2000$. At our signal frequency (shown by the arrow) the noise is predominantly thermal. Reproduceable irregularities in the turntable rotation rate produce peaks at integer values of $f/f_{\rm tt}$. Spurious peaks at odd integer values were suppressed by combining data
taken with 2 opposite orientations of the pendulum in the rotating apparatus.
}
\label{fig: fft}
\end{figure}

The overall performance of the rotating balance is shown in Fig.~\ref{fig: fft}. Except for signals at integer multiples of the turntable frequency caused by reproduceable irregularities in the turntable drive system,  the noise in the twist signal is close to the thermal value expected from internal losses in the tungsten suspension fiber. Systematic errors associated with the turntable signals were eliminated by combining data taken with different orientations of the pendulum within the rotating apparatus.
%
%	Figure 2
%
\subsection{Spin pendulum}
\subsubsection{Design principles}
\begin{figure}[t]
\hfil\scalebox{.33}{\includegraphics*[0.9in,2.6in][7.8in,10.0in]{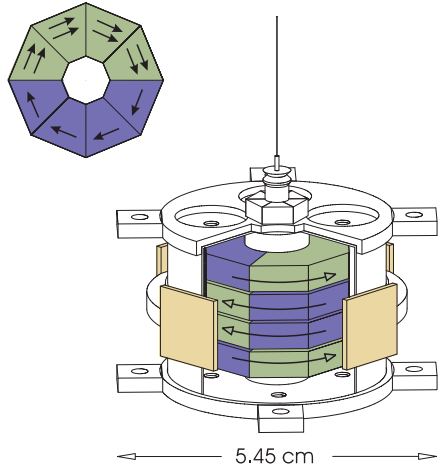}}\hfil
\caption{(color online) Scale drawing of the spin pendulum. The light green and darker blue volumes are Alnico and Sm$\,$Co$_5$, respectively. Upper left: top view of a single ``puck''; the spin moment points to the right. Lower right: the assembled pendulum with the magnetic shield shown cut away to reveal the 4 pucks inside. Two of the 4 mirrors (light gold) used to monitor the pendulum twist are prominent. Arrows with filled heads show the relative densities and directions of the electron spins, open-headed arrows show the directions of ${\bm B}$. The 8 tabs on the shield held small screws that we used to tune out the pendulum's residual $q_{21}$ gravitational moment.}
\label{fig: spin pendulum}
\end{figure}

The heart of our apparatus is a spin pendulum, shown in Fig.~\ref{fig: spin pendulum}, that contains a substantial number of polarized electrons while having a negligible external magnetic moment and high gravitational symmetry. 
The spin pendulum is constructed from 4 octagonal ``pucks''. One half of each puck is Alnico 5 (a conventional, relatively ``soft'' ferromagnet in which the magnetic field is created almost entirely by electron spins) and the other half is grade 22 Sm$\,$Co$_5$ (a ``hard'' rare-earth magnet in which the orbital magnetic moment of the electrons in the Sm$^{3+}$ ion nearly cancels their spin moment\cite{ti:99,gi:79,ko:97}). 

The trapezoidal elements of the pucks were fabricated by electric-discharge machining and kept in precise alignment by aluminum frames. Thin plates glued to the sides of the Alnico pieces compensated for their lower density, $\rho=7.37$~g/cm$^3$ versus 8.3 ~g/cm$^3$ for Sm$\,$Co$_{5}$. After each puck was assembled, we magnetized the Alnico to the same degree as the Sm$\,$Co$_{5}$
by monitoring the external $B$ fields as appropriate
current pulses were sent through coils temporarily wound around the pucks.
By stacking 4 such pucks as shown in Fig.~\ref{fig: spin pendulum}, we placed the effective center of the spin dipole in the middle of the pendulum, reduced systematic magnetic-flux leakage, averaged out the small density differences between Alnico and Sm$\,$Co$_{5}$, and canceled any composition dipole that would have made us sensitive to violation of the weak Equivalence Principle. 

The pucks were surrounded by a gold-coated mu-metal shield that supported 4 mirrors equally spaced around the azimuth. 
The 107 g pendulum was suspended at the end of a 75 cm long tungsten fiber which in turn hung from a mode spoiler that damped the pendulum's ``wobble'', ``swing'' and ``bounce'' oscillations but had essentially no effect on the torsional mode. During the course of this experiment 4 different
fibers were employed with diameters of either $28~\mu$m or $30~\mu$m. The upper attachment of the suspension fiber could be rotated to place any of the 
pendulum's 4 mirrors in the beam of the autocollimator system that monitored the pendulum's twist.

We measured $B_0$, the field inside the pucks, by arranging 11 of the trapezoidal Sm$\,$Co$_{5}$ elements in straight line, and
used an induction coil to determine the field inside the Sm$\,$Co$_{5}$. 
We found $B_0=9.6\pm 0.2 $ kG which was consistent with the supplier's\cite{SmCo5} specification of 9.85 kG.  

\begin{figure}[t]
\hfil\scalebox{.46}{\includegraphics*[46pt,37pt][570pt,448pt]{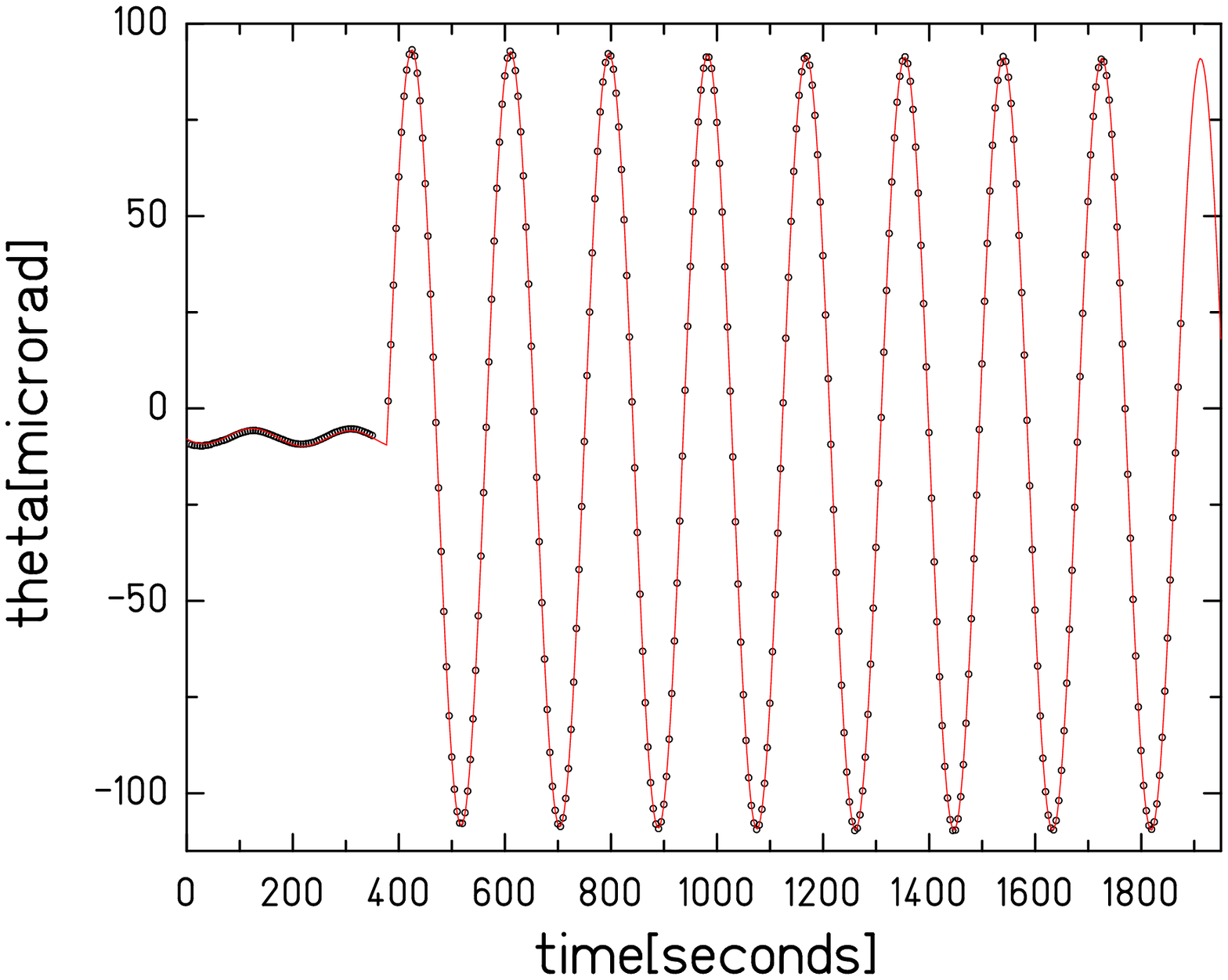}}\hfil
\caption{Dynamical calibration of the torque scale, showing the pendulum twist as a function of time. At $t=377.7$~s the turntable rotation frequency was abruptly changed from
0.55556 mHz to 0.55611 mHz. The resulting change in the pendulum's oscillation amplitude and phase calibrated the angular deflection scale and determined the pendulum's free oscillation frequency $f_0$. The smooth curve is the best fit.}
\label{fig: caleot}
\end{figure}

The absolute calibration of the twist to torque conversion was done by observing the pendulum's response to an abrupt
change in the turntable rotation rate as shown in Fig.~\ref{fig: caleot}. The pendulum's free torsional oscillation frequency, $f_0=5.379$ mHz for this
 $30~\mu$m diameter fiber, together with its calculated rotational inertia, determined the fiber's torsional constant $\kappa=0.185$ dyne-cm/radian. 

%
%	Figure 3
%
%

%
%	Figure 4
%
\subsubsection{Estimating the spin content}
\label{sec: calc of spin content}
Polarized neutron elastic scattering studies\cite{gi:79} on Sm$\,$Co$_5$ have shown that the room-temperature magnetic moment of the Sm$^{3+}$ ion is extremely small, $-0.04\;\mu_{B}$ compared to the $-8.97\;\mu_{B}$ moment of the five cobalts (see Appendix
\ref{ap: spin content}). It is therefore an excellent approximation that the magnetization of Sm$\,$Co$_{5}$ is due entirely to the Co. If we make the additional approximation (which we relax below) that the Co and Alnico magnetizations arise entirely from spins, then
the net number of polarized spins in our pendulum is simply 
\begin{equation}
N_{\rm p}=\frac{B_0 |R|}{\mu_0 \mu_B} V \eta
\label{eq: Npol 0th approx}
\end{equation}
where $B_0$ is the magnetic field inside a puck, $R$ is the ratio of Sm to Co spin moments in room-temperature Sm$\,$Co$_5$ (a negative quantity because the Sm and Co spins are directed oppositely), $\eta=0.65$ accounts for the puck's octagonal shapes and $V=4.12$ cm$^3$ is the total volume of Sm$\,$Co$_{5}$. As discussed in Appendix \ref{ap: spin content}, we used the Sm ion wavefunctions deduced from elastic neutron scattering\cite{gi:79} on Sm$\,$Co$_{5}$ to predict that at
room-temperature the spin contribution to the Sm magnetic moment in Sm$\,$Co$_{5}$ is $\mu_{\rm S}=+3.56\mu_{B}$. Using data from this and other rare-earth/cobalt alloys, we estimate
that the 5 Co's in Sm$\,$Co$_{5}$ have  $\mu_{\rm S}=-7.25 \mu_{B}$ and obtain 
%begin{equation}
$R=-0.49$.
%\end{equation}

On the other hand, a Compton scattering experiment on Sm$\,$Co$_5$ with circularly polarized synchrotron radiation\cite{ko:97}, which directly probes the spin distribution rather than the magnetic field, gives $R=-0.23$. We assume that $R$ is equally likely to lie anywhere in the interval between $-0.49$ and $-0.23$ with a mean value and equivalent Gaussian spread
of 
\begin{equation}
R=-0.36\pm 0.075~.
\end{equation} 
which implies $N_{\rm p}=(0.79 \pm 0.17)\times10^{23}$.

However, as discussed in Appendix \ref{ap: spin content}, experiments have shown that the magnetization of Alnico and of Co in Sm$\,$Co$_{5}$ 
is not entirely due to spins as we assumed above. Defining $f$ as the fractional contribution of spin to the total magnetic moment, we obtain an improved estimate 
\begin{equation}
N_{\rm p}=\frac{B_0 V \eta}{\mu_0 \mu_B}\left[f_{\rm Co}(|R|-1)+f_{\rm Alnico}\right]  \nonumber
\end{equation}
\begin{equation}
=(0.96 \pm 0.17)\times 10^{23}~~~~~~~
\label{eq: Npol}
\end{equation}
%xxx program npol xxx
where $f_{\rm Alnico}=0.953\pm 0.005$ and $f_{\rm Co}=0.81\pm 0.03$ are taken from Eqs. \ref{eq: f Alnico} and \ref{eq: f Co} below.
 
We show in Sec.~\ref{sec: Np calib} below that the Earth's rotation acting on the pendulum's angular momentum provided an internal calibration of $N_{\rm p}$ in very good agreement with the value in Eq.~\ref{eq: Npol}.
\section{Experimental signatures}
\label{sec: signatures}
\subsection{General form of the signatures}
The potentials of interest (Eqs. \ref{eq: def A}, \ref{eq: def B}, \ref{eq: def C}, \ref{eq: non-com pot}, \ref{eq: MW} and \ref{eq: BD})  all imply that the pendulum's energy has the form
\begin{equation} 
E=-N_{\text p}\:{\bm\sigma}_{\text p} \cdot \bm{\beta}~,
\end{equation} 
where $\bm{\sigma}_{\rm p}$ represents the orientation of the pendulum's spin, and $\bm{\beta}$ has a value in the laboratory frame that may depend on time. This potential applies a torque
$\bm{\tau}=N_{\rm p}\:\bm{\sigma}_{\rm p} \bm{\times \beta}$ on the pendulum that can be detected by measuring its induced twist. However, only vertical torques, which are resisted by the very soft torsion spring constant, $\kappa$, produce a measurable twist; torques in the horizontal plane are resisted by the vastly stronger ``gravitational spring" with an effective constant
$\kappa_{\rm grav}= Mgs\approx 2.4\times 10^6~\kappa$ where $M$ is the pendulum mass and $s=2.72$ cm is the vertical distance from the pendulum's center of mass to the fiber attachment point. As a result, 
the pendulum's spin is confined to the local horizontal plane and rotates along with the turntable about a vertical axis. Although the pendulum does not exactly follow the turntable angle because external torques on the pendulum twist the suspension fiber by angles $\theta \alt 10^{-6}$ radians, to an excellent approximation, the pendulum twist, $\theta$, as a function of turntable angle, $\phi$, has the form
\begin{eqnarray}
\theta(\phi)&=& (\beta_{\perp} N_{\rm p}/\kappa) \sin(\phi_0-\phi) \nonumber \\
 &\equiv& -(N_{\rm p}/\kappa)(\beta_{\rm N}\sin\phi+\beta_{\rm E}\cos \phi)~.
\end{eqnarray}
Here $\beta_{\perp}$ is the projection of $\bm{\beta}$ on the horizontal plane, $\phi_0$ its azimuthal angle, and $\beta_{\rm N}$ and $\beta_{\rm E}$ are the North and East components of $\bm{\beta}_{\perp}$. We use the convention that all angles are measured counterclockwise as seen from above, and that $\phi=0$ points North. The procedure for converting our twist measurements into values for $\beta_{\rm N}$ and $\beta_{\rm E}$ is discussed in Section~\ref{sec: extracting} below.

The scenarios we test in this paper have signatures where $\bm{\beta}$ is fixed in the laboratory (Eq.~\ref{eq: MW}), and others where $\bm{\beta}$ is fixed in inertial space (Eqs.~\ref{eq: def A} and \ref{eq: non-com pot})
or depends on the orientation and/or velocity of the earth's motion around the sun (Eqs.~\ref{eq: def B}, \ref{eq: def C} and \ref{eq: BD}). These time-varying signatures were computed using astronomical formulae
given by Meeus.\cite{me:98} We neglected the earth's radius and the surface velocity associated with its spin in comparison to the astronomical unit and the earth's orbital velocity, respectively.
\subsection{Lorentz-symmetry violation}
We constrained the rotational symmetry violating parameters of Eqs.~\ref{eq: def A} and \ref{eq: non-com pot} by searching for a
$\bm{\beta}$ fixed in inertial space.
Even though we are sensitive only to the two horizontal components of $\bm{\beta}$, 
the earth's rotation allowed us to probe all three components in the inertial frame,
\begin{eqnarray}
\beta_{\rm N}(t)&=&[A_X\;T^{\rm N}_X(t) + A_Y\;T^{\rm N}_Y(t)+A_Z\;T^{\rm N}_Z(t)] \nonumber \\
                &=&-(A_X\cos \Omega t + A_Y\sin \Omega t)\sin \Psi  + A_Z \cos\Psi  \nonumber \\
\beta_{\rm E}(t)&=&[ A_X\;T^{\rm E}_X(t) + A_Y\;T^{\rm E}_Y(t)+A_Z\;T^{\rm E}_Z(t)] \nonumber \\
                &=& -A_X \sin \Omega t + A_Y\cos \Omega t ~.
\label{eq: sidereal}
\end{eqnarray}
The $T^{\rm N,E}_{X,Y,Z}(t)$ coefficients project the equatorial $X,Y,Z$ system onto local N,E horizontal coordinates;  
$\Psi=47.658$ deg is the latitude of our laboratory, $\Omega=2\pi/0.9972696$d is the sidereal frequency and $t$ is the local apparent sidereal time.

\begin{figure}
\hfil\scalebox{.53}{\includegraphics*[39pt,34pt][483pt,654pt]{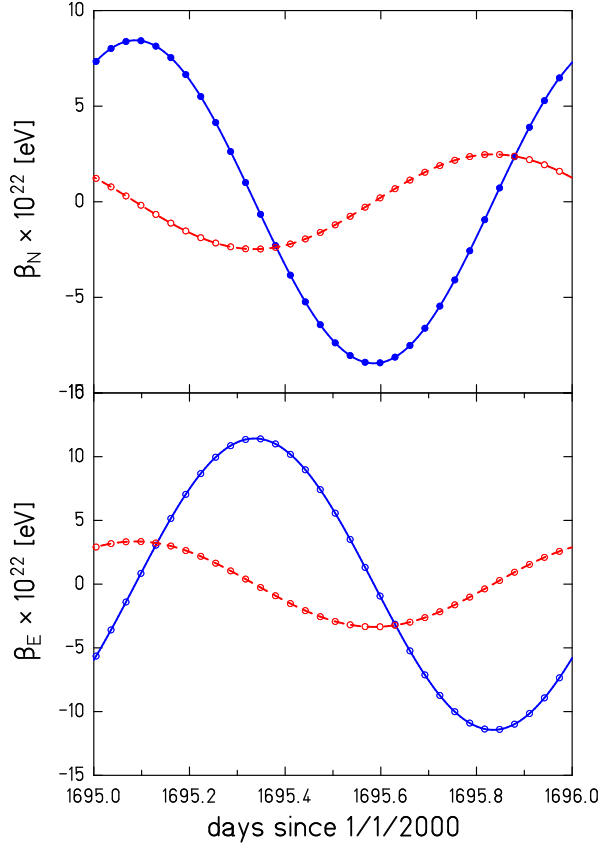}}\hfil
\caption{(color online) Daily signals produced by the Lorentz-boost violating term $C_{ij}$. The daily signals can be essentially constant (i.e. with a weak annual modulation), modulated at the sidereal frequency, or modulated at both the sidereal and annual frequencies . Solid blue line: $C_{XX}=10^{-18}$~eV; dashed red line $C_{YY}= 10^{-18}$~eV. The points show the actual times at which data were acquired.}
\label{fig: boost signals}
\end{figure}

\begin{figure}
\hfil\scalebox{.6}{\includegraphics*[57pt,35pt][477pt,378pt]{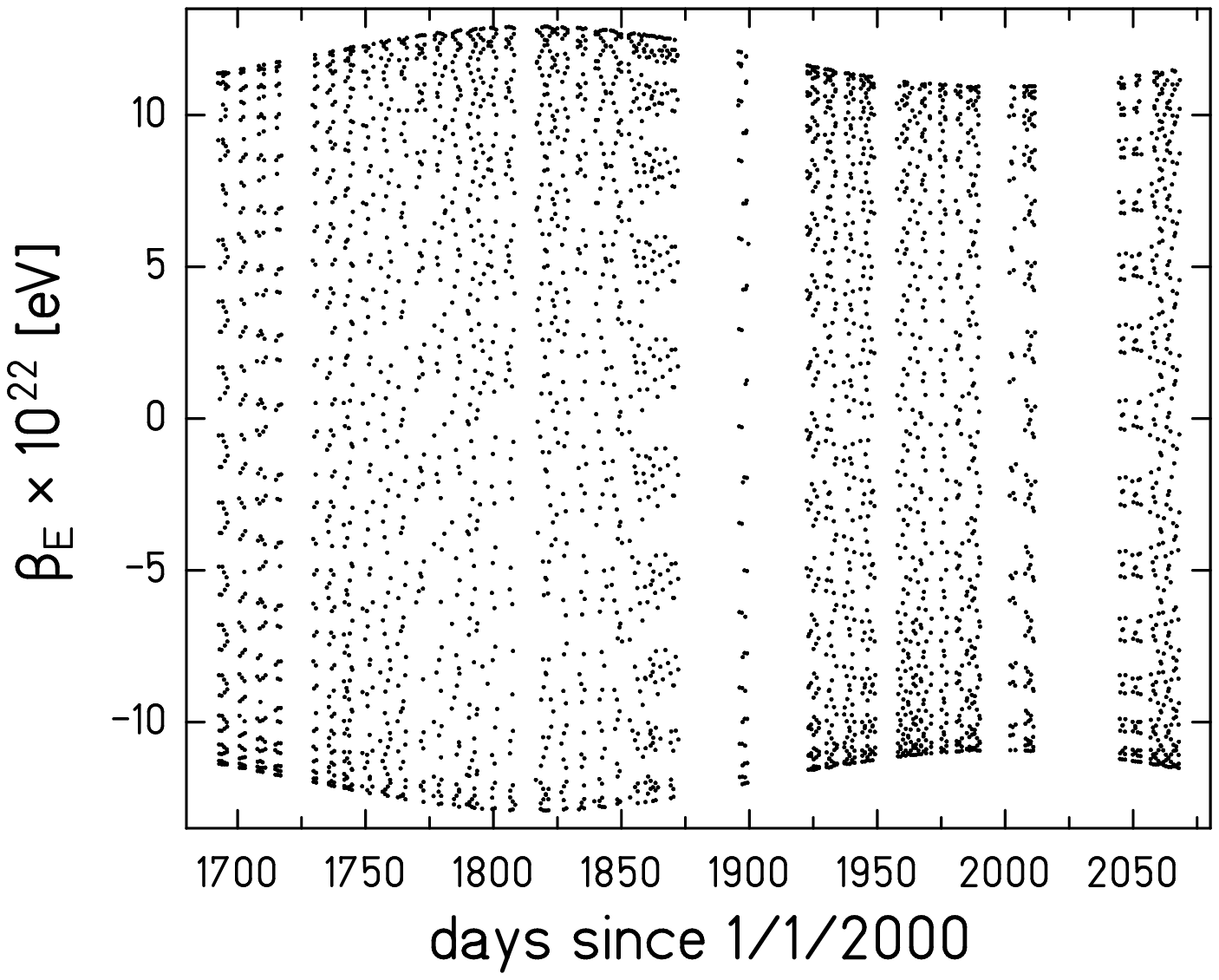}}\hfil
\caption{Expected $\beta_{\rm E}$ signal corresponding to $C_{XX}=10^{-18}$~eV.  Each point is placed at the time of one of the cuts in our complete data set. Almost 4000 additional data points extending out to 2825 on the horizontal axis are omitted to keep the figure readable. The annual modulation arises from $\bm{v}_{\odot}$, the steady component comes from $\bm{v}_{\rm CMB}$.}
\label{fig: XX boost signal}
\end{figure}

We measured the boost parameters in Eqs.~\ref{eq: def B} and \ref{eq: def C} 
using the motion of the earth with respect to the entire cosmos (neglecting the much smaller velocity of our laboratory with respect to the center of the earth). The velocity of the earth with respect to the rest frame of the cosmic microwave background (CMB) is
\begin{equation}
\bm{v}_{\oplus}=\bm{v}_{\odot}+\bm{v}_{\rm CMB}~,
\end{equation}
where $\bm{v}_{\odot}$ is the velocity of the earth with respect to the sun and 
$\bm{v}_{\rm CMB}$
is the velocity of the sun in the CMB rest frame. In equatorial coordinates 
\begin{eqnarray}
\bm{v}_{\rm CMB}&=& c(-1.190\bm{\hat{X}}+ 0.253\bm{\hat{Y}} -0.165\bm{\hat{Z}})10^{-3} \\ 
\bm{v}_{\odot}&=& v_{\odot}^x \bm{\hat{X}} + v_{\odot}^y( \cos\epsilon\bm{\hat{Y}} + \sin\epsilon \bm{\hat{Z}})~,
\end{eqnarray}
where $\epsilon=23.45$ degrees is the inclination of the ecliptic.
The ecliptic coordinate $\bm{z}$ is perpendicular to the plane of the earth's orbit, $\bm{x}=\bm{X}$, and $\bm{y}=
\bm{z} \times \bm{x}$ so that
\begin{eqnarray} 
v_{\odot}^x &=& -{\dot R_{\odot}} \cos\lambda_{\odot}+\dot{\lambda}_{\odot} R_{\odot} \sin\lambda_{\odot}\:  \nonumber \\
v_{\odot}^y &=& -{\dot R_{\odot}} \sin\lambda_{\odot}-\dot{\lambda}_{\odot} R_{\odot} \cos\lambda_{\odot}\: 
\label{eq: velocities}
\end{eqnarray} 
where $\lambda_{\odot}(t)$, the true longitude of the sun, and $R_{\odot}(t)$, the geocentric distance to the sun, were computed using expressions given by Meeus\cite{me:98}. Converting 
$\bm{v}_{\odot}$ from ecliptic to equatorial coordinates, we have
\begin{equation}
\beta_{\rm N,E}(t)=B\sum_i  \, \frac{v_{\oplus}^i(t)}{c}\; T^{\rm N,E}_i(t)~,
\label{eq: B signal}
\end{equation}
and
\begin{equation}
\beta_{\rm N,E}(t)=\sum_{i,j} C_{ij}\, \, \frac{v_{\oplus}^j (t)}{c}\; T^{\rm N,E}_i(t)~.
\label{eq: C signal}
\end{equation}
Figure~\ref{fig: boost signals} shows the signatures of boost-symmetry violating terms with $C_{XX}=10^{-18}$~eV and with 
$C_{XZ}=10^{-17}$~eV.  Figure~\ref{fig: XX boost signal} shows how the $\beta_{\rm E}$ signature varies over the course of a year.
The boost-violating signatures, which are modulated at the sidereal, annual and annual times sidereal frequencies, can be distinguished from a violation of rotational-invariance which have only a sidereal modulation. Note that the signatures of $A_Z$ as well as $C_{ZX}$, $C_{ZY}$ and $C_{ZZ}$ do not have daily modulations and must be extracted from lab-fixed $\beta_{\rm N}$'s. 

%	Figure 5
%
%

%	Figure 6
%
\subsection{Exotic boson-exchange interactions}
\subsubsection{Static potentials}
\begin{figure}[t]
\hfil\scalebox{.7}{\includegraphics*[51pt,31pt][360pt,470pt]{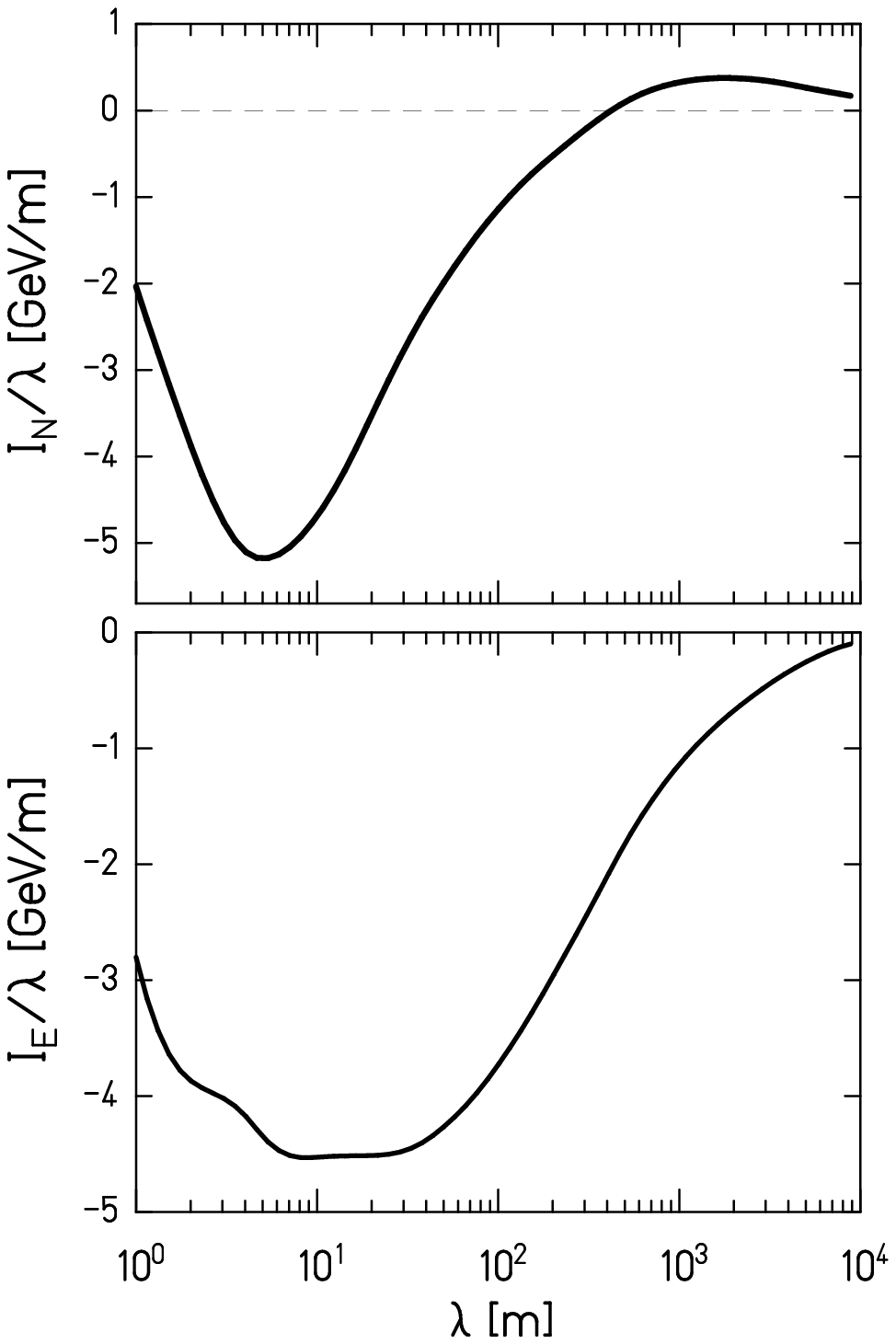}}\hfil
\caption{Earth source integrals defined in Eq.~\ref{eq: earth source}. The vertical axis shows $I(\lambda)$ divided by $\lambda$.
To obtain constraints on $g_{\rm P}^e g_{\rm P}^e$ or $g_{\rm P}^e g_{\rm P}^p$, rather than 
on $g_{\rm P}^e g_{\rm P}^N$, simply multiply
$I(\lambda)$ by $[Z/\mu]_{\oplus}\approx 0.48$.}
\label{fig: source int}
\end{figure}

A $CP$-violating monopole-dipole interaction (Eq.~\ref{eq: MW}) between the spin pendulum
and the earth would give time-independent $\beta_{\rm N}$ and  $\beta_{\rm E}$ values that depend upon the range, $\lambda$, of the interaction. 
\begin{equation}
\beta_{\rm N,E}=-\frac{g_{\rm P}^e g_{\rm P}^N}{\hbar c} I_{\rm N,E}(\lambda)~,
\end{equation} 
where 
\begin{equation}
I_{\rm N,E}(\lambda) = \bm{\hat{e}}_{\rm N,E} \bm{\cdot}\!\! \int\! \bm{\hat{r}} \left[ \left( \frac{1}{r \lambda}+\frac{1}{r^2}\right)e^{-r/\lambda}\right]\frac{K\rho(\bm{r})}{u}d^3\bm{r}~, 
\label{eq: earth source}
\end{equation}
where $\bm{r}$ points from the pendulum to a source mass element with density $\rho$, $\bm{\hat{e}}_{\rm N,E}$ is a unit vector pointing N or E, 
\begin{equation}
K=\frac{\hbar^2}{8\pi m_e }~,
\end{equation}
and $u$ is the atomic mass unit. 
This integral is, to within a constant factor, identical to that needed to evaluate the E\"ot-Wash Equivalence Principle results\cite{ad:90,su:94}, so we adopted the source integrations of those works. 
As explained in Refs.~\cite{ad:90,su:94},
the integral can be reliably computed in two regimes: 1 m $\leq \lambda \leq 10$~km (where information from local topography and boreholes is sufficient, and $\lambda \geq 1000$~km where smooth, layered models of the entire earth are adequate. It is difficult to evaluate the integral in the gap region, and we have not done so. The integrals are displayed in Fig.~\ref{fig: source int}.
%
%	Figure 7
%
\subsubsection{Interactions between the spin pendulum and the sun}
A $CP$-violating monopole-dipole interaction (Eq.~\ref{eq: MW}) between the spins in our pendulum
and the sun would give modulated $\beta_{\rm N}$ and  $\beta_{\rm E}$ values
\begin{eqnarray}
\beta_{\rm N}(t)\!&=&\! -(g_{\rm P}^e g_{\rm P}^N/\hbar c) I_{\odot}(\lambda) \cos\alpha_{\odot}(t) \cos\gamma_{\odot}(t)/R_{\odot}^2(t)~ \nonumber \\
\beta_{\rm E}(t)\!&=&\! -(g_{\rm P}^e g_{\rm P}^N/\hbar c) I_{\odot}(\lambda)\cos\alpha_{\odot}(t) \sin \gamma_{\odot}(t)/R_{\odot}^2(t)~, \nonumber \\
 & & 
\label{eq: CP sun}
\end{eqnarray} 
where $\alpha_{\odot}$ and $\gamma_{\odot}$ are the altitude and azimuth (measured Eastward from North) of the sun. 
For a uniformly dense sun of mass $M_{\odot}$ and radius $r_{\odot}$, the source integral becomes\cite{ni:87}
\begin{equation}
I_{\odot}(R_{\odot},\lambda)\!=  \frac{K M_{\odot}}{u} \! \left[\frac{\xi\cosh \xi \!- \!\sinh \xi}{\xi^3/3}\right]\!\! \left[ \frac{R_{\odot}}{\lambda}\!+\!1 \right] \exp\left[-\frac{R_{\odot}}{\lambda}\right]~,
\label{eq: source 1}
\end{equation}
where $\xi=r_{\odot}/\lambda$.

We searched for the long-range velocity-dependent boson-exchange interactions of 
Eq.~\ref{eq: BD} using the earth's velocity around the sun. We needed, in addition to the velocity given in Eq.~\ref{eq: velocities} and $I_{\odot}(R,\lambda)$, the term $\bm{\tilde{v}} \times  \bm{\hat{r}}=-\dot{\lambda}_{\odot} R_{\odot}/c\:\hat{\bm e}_z$, and the source integral 
\begin{equation}
I_{\odot}^v(R_{\odot},\lambda)=  \frac{M_{\odot}}{8\pi u}\left[ \frac{\xi\cosh \xi - \sinh \xi}{\xi^3/3}\right] \exp\left[-\frac{R_{\odot}}{\lambda}\right]~.
\label{eq: source 2}
\end{equation}
Then 
\begin{eqnarray}
\label{eq: BD sun}
\beta_{\rm N}(t)\! &=&\! f_{\perp} I_{\odot}{\dot\lambda}_{\odot}\:(\cos\alpha_z \cos{\gamma_z})/(\hbar c^2R_{\odot})\\
&-&\! f_v I^v_{\odot} \:(\tilde{v}_x \cos \alpha_x \cos \gamma_x + \tilde{v}_y\cos \alpha_y \cos \gamma_y)/R_{\odot}  \nonumber\\
\beta_{\rm E}(t) &=& f_{\perp} I_{\odot}{\dot\lambda}_{\odot}\: (\cos\alpha_z \sin{\gamma_z}) /(\hbar c^2R_{\odot}) \nonumber \\
&-& f_v I^v_{\odot} \:(\tilde{v}_x \cos \alpha_x \sin \gamma_x + \tilde{v}_y\cos \alpha_y \sin \gamma_y) /R_{\odot} \nonumber
\end{eqnarray}
where $\alpha(t)_{x(y,z)}$ and $\gamma(t)_{x(y,z)}$ are the laboratory altitude and azimuth of the ecliptic unit vectors $\hat{x}$, 
$\hat{y}$ and $\hat{z}$.
When evaluating the $\lambda$-dependence of the couplings $g_{\rm P}^e g_{\rm S}^A$, $f_v$ and $f_{\perp}$  of Eqs.~\ref{eq: MW} and \ref{eq: BD}, we approximated the factors $R$ in Eqs.~\ref{eq: source 1}-\ref{eq: source 2}
by their average value $R=1$ AU. 
\subsubsection{Interactions between the spin pendulum and the moon}
We improved our $f_v$ and $f_{\perp}$ constraints for $\lambda << 1$~AU by using the moon's motion about the earth. The moon's coordinates and velocity were evaluated using expressions from Chapter 47 of Meeus\cite{me:98} to compute generalizations of Eqs.~\ref{eq: BD sun} which are more complicated because the moon's motion does not lie in the ecliptic plane and is strongly perturbed by the gradients in the sun's field.
\subsection{Gyro-compass effect}
\label{sec: gyro}
Because the pendulum's magnetic flux was confined entirely within the pucks, the stationary pendulum necessarily contained a net total angular momentum $\bm{J}=-\bm{S}$, where $S=N_{\rm p} \hbar/2$ is the pendulum's net spin. The magnetic fields in the
two materials,
\begin{eqnarray}
\!\!\!B_z({\rm SmCo}_5)/\mu_B\!&=&\!g_s S_z({\rm SmCo}_5)+g_l L_z({\rm SmCo}_5) \\
\!\!\!B_z({\rm Alnico})/\mu_B\!&=&\!g_s S_z({\rm Alnico})+g_l L_z({\rm Alnico})~, \nonumber 
\end{eqnarray}
were equal and opposite so that the net magnetic field in the pendulum's $z$ direction vanished. This requires that
\begin{equation}
2 \langle S_z^{\rm tot} \rangle+\langle L_z^{\rm tot}\rangle=0~,
\end{equation}
where we make the excellent approximation that $g_s=2$.
This implies
\begin{equation}
\langle J_z^{\rm tot} \rangle=-\langle S_z^{\rm tot}\rangle~.
\end{equation}

To keep $\bm{J}$ fixed in the frame of the rotating earth, the fiber has to apply a steady torque, $\bm{T}$, given by
\begin{equation}
\bm{T \cdot \hat{n}}=|\bm{\Omega_{\oplus} \times J \cdot \hat{n}}|~,
\label{eq: gyro}
\end{equation} 
where  $\bm{\Omega_{\oplus}}$ is sidereal rotation frequency and $\bm{\hat{n}}$ is the local vertical. The pendulum twists until $\kappa \theta=-\bm{T \cdot \hat{n}}$, i.e. $\bm{J}$ tries to point toward true North with a torque
\begin{equation}
\bm{T\cdot \hat{n}}=N_{\rm p}\frac{\hbar\Omega_{\oplus}\cos{\Psi}}{2}=N_{\rm p}\times(2.590\times 10^{-39}\:{\rm Nm})~,
\label{eq: gyro torque}
\end{equation}
where $\Psi$ is defined in Eq.~\ref{eq: sidereal}.
This is equivalent to a
small negative (because $\bm{J}=-\bm{S}$) value
\begin{equation}
\beta_{\rm N}^{\rm gyro}=-1.616\times 10^{-20}~{\rm eV}~,
\label{eq: gyro beta}
\end{equation}
which formed a steady, precisely known background in our study of lab-fixed signals, and allowed us to calibrate the spin content of our pendulum.  
\section{Measurements}
\label{sec: measurements}
The data for this paper were taken in two distinct sets. Runs in set I were taken over a span of 18 months
between August 2004 and January 2006. Data were taken with the autocollimator beam reflected by each of the pendulum's 4 mirrors. During this period, along with acquiring normal spin-pendulum data, we made extensive studies of systematic effects from gravity-gradient and magnetic couplings and from turntable imperfections.
Despite considerable effort, that included changing suspension fibers and mode spoilers, we were unable to understand a slowly-varying systematic effect, and as a result our laboratory-fixed constraints were dominated by poorly understood systematics.   
This work formed the basis of our brief publication\cite{he:06}. 

\begin{figure}[t]
\hfil\scalebox{.5}{\includegraphics*[100pt,231pt][400pt,567pt]{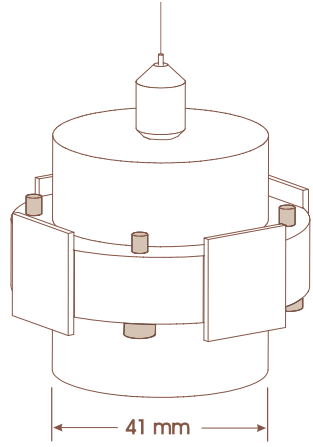}}\hfil
\caption{Scale drawing of the ``zero-moment'' pendulum. The pendulum body was made from 7075 aluminum. The 4 vertical screws (shaded) were adjusted to trim away the residual $q_{21}$ moments of the pendulum. The ``ball-cone'' device on top of the pendulum allowed us to rotate the pendulum with respect to the torsion fiber. The entire pendulum, including the mirrors, was coated with gold.}
\label{fig: zero moment}
\end{figure}

We then decided to tackle the problem of slowly varying systematics by constructing the 93 g ``zero-moment'' pendulum shown in Fig.~\ref{fig: zero moment}. This aluminum pendulum had no net spin, was nonmagnetic and had the same moment of inertia as the spin pendulum. It coupled negligibly to gravity gradients because it had cylindrical symmetry and very small $q_{20}$, $q_{30}$ and $q_{40}$ mass moments. In addition, we modified the
fiber attachment to the pendulum so that we could reproduceably rotate the pendulum with respect to its suspension system. 
As discussed below, this device revealed an unexpected systematic effect associated with the suspension system that could be canceled by combining results taken with opposite orientations of the pendulum with respect to the suspension system.
% as shown in Fig.~\ref{fig: ballcone}. 

We added the new rotatable fiber attachment mechanism to the spin pendulum, and set II runs occured  between September 2006 and March 2008. Data were taken with the light beam reflecting from each of the 4 mirrors and for 4 equally spaced angles between the pendulum and the suspension system. The results from set II gave much tighter lab-fixed constraints, and when combined with the set I data improved the sidereal and solar signals published in Ref.~\cite{he:06}.
%
%	Figure 8
%

\section{Data analysis}
\label{sec: analysis}
\subsection{Extracting $\bm{\beta_{\rm N}}$ and $\bm{\beta_{\rm E}}$ from the pendulum twist}
\label{sec: extracting}
Over the course of this experiment the turntable rotation frequency, $f$, was set at values between $3f_0/29$ and $3f_0/20$, where $f_0$ is the free-oscillation frequency of the pendulum. 
We recorded the pendulum twist angle $\theta$ as a function of the angle $\phi$ of the turntable (measured counter clock-wise from North), and converted it to torque, as described in Ref.~\cite{su:94}. Briefly, data were divided into ``cuts'' each of which contained exactly 2 revolutions of the turntable, lasted no more than 3800 s and typically contained 254 data points. A simple digital filter was applied to suppress the free oscillations. The twist data from each cut were then fitted by 2nd-order polynomial ``drift'' terms (typical drift was $0.5~\mu$rad/h) and harmonic terms 
\begin{equation}
\theta(\phi)=\sum_{n=1}^8 (a^n_{\rm N} \cos n\phi -  a^n_{\rm E} \sin n \phi)~.
\end{equation}
Our signal was contained in the  
$n=1$ coefficients that we denote below as $a_{\rm N}(\phi_{\rm d},\phi_{\rm s})$ and $a_{\rm E}(\phi_{\rm d},\phi_{\rm s})$, where $\phi_{\rm d}$ is the
angle of the pendulum (normally its spin dipole) with respect to the turntable and $\phi_{\rm s}$ is the angle between the pendulum and its suspension
system.

We converted the $a_{\rm N}$ and $a_{\rm E}$ signals for each cut into 
$\beta_{\rm N}$ and $\beta_{\rm E}$ values using

\begin{eqnarray}
\left[\! \begin{array}{r} a_{\rm N}(\phi_{\rm d},\phi_{\rm s}) \\ a_{\rm E}(\phi_{\rm d},\phi_{\rm s})\end{array}\! \right]=
\left[\! \begin{array}{rr}\cos \phi_{\rm d} & \sin \phi_{\rm d} \\ -\sin \phi_{\rm d} & \cos \phi_{\rm d} \end{array}\!\right]\!\! \left[\! \begin{array}{r}  \delta_{\rm N} \\ \delta_{\rm E} \end{array}\!\right] + \left[\! \begin{array}{r} \delta_{\rm N}^{\rm t}(\phi_{s}) \\\delta_{\rm E}^{\rm t}(\phi_{s}) \end{array}\!\right] + \nonumber \\
 \left[\! \begin{array}{rr} \cos(\phi_{\rm d}\!+\!\phi_{\rm s}) & \sin(\phi_{\rm d}\!+\!\phi_{\rm s}) \\ -\sin(\phi_{\rm d}\!+\!\phi_{\rm s}) & \cos(\phi_{\rm d}\!+\!\phi_{\rm s})\end{array}\!\right]\!\!\left[\! \begin{array}{r} \delta_{\rm N}^{\rm s} \\ \delta_{\rm E}^{\rm s}\end{array}\!\right]~.
\label{eq: lab-fixed decomp}
\end{eqnarray}
The fit parameters ($\delta_{\rm N}$, $\delta_{\rm E}$) are the twists associated with the pendulum itself, so that
\begin{equation}
\beta_{\rm N,E}=\kappa \delta_{\rm N,E}/N_{\rm p}~,
\label{eq: betas}
\end{equation}
while ($\delta_{\rm N}^{\rm s}$, $\delta_{\rm E}^{\rm s}$)
and ($\delta_{\rm N}^{\rm t}$, $\delta_{\rm E}^{\rm t}$)
are spurious twists associated with the orientations of the suspension system and turntable rotation-rate irregularities, respectively. For data taken with the $30\mu$m fiber, $\delta_{\rm N,E}=1$~nrad corresponds to $\beta_{\rm N,E}=
1.2 \times 10^{-21}$ eV.
\subsection{Astronomically modulated signals}
%
%	Figure 9
%
\begin{figure}[t]
\hfil\scalebox{.59}{\includegraphics*[61pt,36pt][454pt,495pt]{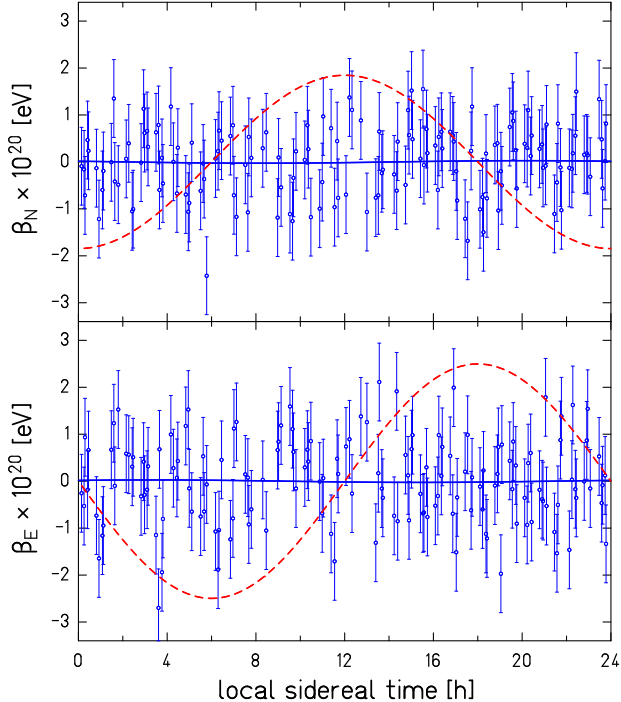}}\hfil
\caption{Data from a set of runs at $\phi_{\rm d}=112.5$ deg that spanned a duration of 113 h. The sidereal average of the signals has been set to zero.
The dashed curves show the signal from a hypothetical ${\bm A}=(2.5 \times 10^{-20}~{\rm eV})\bm{\hat{x}}$ which gives out-of-phase sine waves in $\beta_N$ and $\beta_E$. The solid curves show the best sidereal fit, which yields $A_x=(-0.20 \pm 0.76)\times 10^{-21}$ eV and $A_y=(-0.23 \pm 0.76)\times 10^{-21}$ eV. (Note that the hypothetical signal in Fig. 2 of Ref.~\cite{he:06} was actually ${\bm A}=(2.5 \times 10^{-20}~{\rm eV})\bm{\hat{x}}$.)}
\label{fig: typical data}
\end{figure}
When analyzing our results for astronomically modulated signals, it was sufficient to ignore the last matrix product term on the right hand side of 
Eq.~\ref{eq: lab-fixed decomp}. 
We adopted the strategy of Ref.~\cite{su:94} and assumed that $\bm{\beta}$'s modulated at daily, sidereal or yearly periods could be treated as essentially constant during any one cut. 
This approximation attenuated sidereal signals by at most 0.3\% and was neglected. We used linear regression to fit these measured $\bm{\beta}$'s with the expected signatures
at the midpoint of each cut. Our input consisted of 93 data sets (with durations ranging from 1 to 5 days) containing 8471 cuts and spanned a total of 1110 days. We minimized our sensitivity to slow drifts in our signals by zeroing the average values of the twist signals in each data set as well as any steady average value of the predicted signatures.
The individual data sets were weighted by the inverse-squares of the spread of the $\beta$'s in that set.  
We constrained $\bm A$ in Eq.~{\ref{eq: def A} by fitting our data with the two free parameters $A_x$ and $A_y$ given in Eq.~\ref{eq: sidereal}.
Figure~\ref{fig: typical data} shows a typical data set and the fit. 
The helicity term in Eq.~\ref{eq: def B} was constrained by fitting Eq.~\ref{eq: B signal} with the single free parameter $B$. 
An analysis with 6 free parameters provided the constraints on the 6 modulated boost-symmetry violating $\bm {C}$ terms in Eqs.~\ref{eq: def C} and \ref{eq: C signal} 
%shown in Table~\ref{tab: Kost d params}; 
(only 6 of the 9 $C_{ij}$'s produce modulated signals because spins pointing along $z$ give a vanishing E signal and a steady N signal). 
Limits on the boson exchange potentials  defined in Eqs.~\ref{eq: MW} and \ref{eq: BD} obtained using the Sun as a source were found by fitting Eqs.~\ref{eq: CP sun} and \ref{eq: BD sun} with the 3 free parameters $g_{\rm P}^eg_{\rm S}^A$, $f_{\perp}$ and $f_v$. 
Because of the inclination of the earth's rotation axis, these
torques have components modulated with a 24 hour period as well as 
annual modulations. 
In every case, the resulting fit had $\chi^2/\nu=0.99$. Because the individual data sets were normalized to $\chi^2/\nu=1$, a $\chi^2/\nu=0.99$ for the combined 93 data sets showed that there was no evidence for any anomalous behavior.
The sidereal or solar modulation of the signatures eliminated many systematic effects that were fixed in the lab; as a consequence our bounds on  $A_X$ and $A_Y$ and on the 6 modulated $C$ terms are tighter than those on $A_Z$ or $C_{ZX}$, $C_{ZY}$ and $C_{ZZ}$ which are based on the lab-fixed limits discussed below (see Tables~\ref{tab: A params} and  \ref{tab: C params} in Sec.~\ref{sec: results}). 
\subsection{Laboratory-fixed signals}
\subsubsection{Data set I}
We analyzed our $\beta_{\rm N}$ and $\beta_{\rm E}$ data  for torques fixed in the lab frame by  combining the signals observed for 4 equally-spaced angles, $\phi_{\rm d}$, of the spin dipole
within the rotating apparatus. As described in Ref.~\cite{he:06}, we observed a scatter that was larger than our statistical uncertainties. This  prompted us to study lab-fixed systematic effects with the ``zero-moment'' pendulum.
\subsubsection{Systematic investigations with the zero-moment pendulum}
%
%
%	Figure 10
%
\begin{figure}[!h]
\hfil\scalebox{.65}{\includegraphics*[56pt,37pt][421pt,359pt]{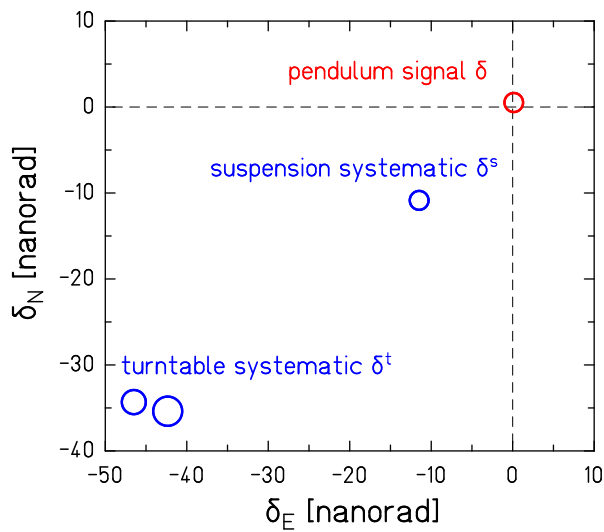}}\hfil
\caption{(color online) Systematic error study using the ``zero-moment'' pendulum.
Data were taken for 4 orthogonal directions of the pendulum with respect to the turntable
and for two opposite orientations of the pendulum with respect to its suspension system and fitted with 
Eq.~\ref{eq: lab-fixed decomp}. The ellipses show the values and $1\sigma$ uncertainties in the extracted signals associated with the pendulum, the turntable and the suspension system, respectively. 
Systematic effects associated with the pendulum itself were negligible.}
\label{fig: ballcone}
\end{figure}
We analyzed data taken on all 4 mirrors and with two opposite orientations of the suspension system with respect to the pendulum using Eq.~\ref{eq: lab-fixed decomp}, where the $\phi_{\rm d}$ was defined by a ficticious zero mark on the pendulum. The results, shown in Fig.~\ref{fig: ballcone},
revealed an unexpected systematic effect associated with the pendulum suspension system, along with the known turntable systematic.
Figure~\ref{fig: ballcone} shows that $\delta_{\rm N}^{\rm t}$ and $\delta_{\rm E}^{\rm t}$ changed slightly every
time we stopped the turntable and opened the vacuum system to change $\phi_{\rm s}$. Therefore those parameters are shown as
functions of $\phi_{\rm s}$ in Eq.~\ref{eq: lab-fixed decomp}. There was no resolvable systematic effect associated with the pendulum. We therefore base all our lab-fixed constraints on the set II data, which did not suffer from the suspension systematic. 

\subsubsection{Data set II}
Spin-pendulum data were acquired with 4 orientations of the pendulum with respect to the rotating apparatus, and for 2 or 4 orthogonal orientations of 
of the suspension system with respect to the pendulum. Three such measurements were taken: Measurement 1 (centered around  October 15, 2006), Measurement 2 (centered around June 15, 2007), and Measurement 3 (centered around March 8, 2008).
We used Eqs.~\ref{eq: lab-fixed decomp} and \ref{eq: betas} to extract from each of these measurement ``true'' $\beta_{\rm N}$ and $\beta_{\rm E}$ values that depended only on the orientation of the spin dipole.
The extracted values of $\beta_{\rm N}$ and $\beta_{\rm E}$ for one of these measurements are 
%listed in Table~\ref{tab: Phase II lab fixed} and 
plotted in Fig.~\ref{fig: Phase II lab fixed}. A highly significant signal was observed (see Table~\ref{tab: Phase II lab fixed} below) that points within $0.2\pm 1.6$ degrees of due South; it agrees well with the expected gyro-compass effects discussed in Sec.~\ref{sec: gyro}.
%
%	Fig 11
%
\begin{figure}
\hfil\scalebox{.75}{\includegraphics*[49pt,30pt][340pt,347pt]{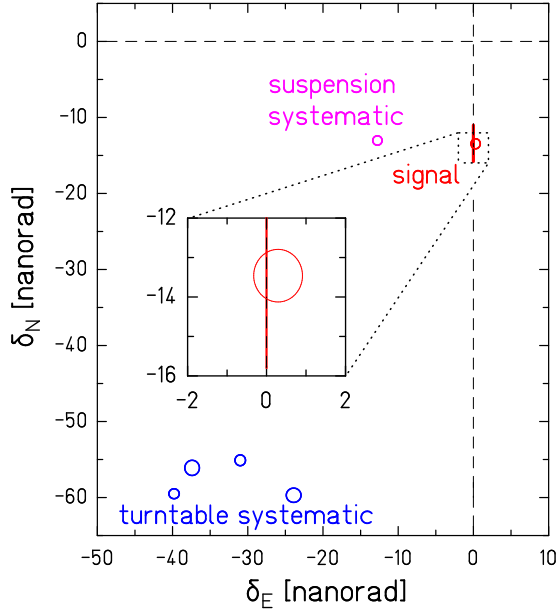}}\hfil
\caption{(color online) Extraction of lab-fixed signals from the October 2006 data, showing the N and E components of the twist signals, obtained using Eq.~\ref{eq: lab-fixed decomp}. Ellipses indicate $1\sigma$ statistical errors.  The turntable systematic changed each time we opened up the vacuum system to change $\phi_{\rm ds}$, while the suspension systematic remained constant at a value very close to that seen in the ``zero-moment'' data
plotted in Fig.~\protect\ref{fig: ballcone}. The red vertical lines shows the expected gyrocompass signal;
the length of the line reflects the uncertainty in the $R$ value.} 
\label{fig: Phase II lab fixed}
\end{figure}
\section{Systematic errors}
\label{sec: systematics}
The interactions in Eqs.~\ref{eq: def A}$-$\ref{eq: BD} all produce 1$\omega$ twist signals. Any other mechanism that can produce a 1$\omega$ signal is a
potential source of systematic error.  We present below systematic error analyses for two types of such 
effects: lab-fixed signals whose amplitude and phase are constant in time and astronomical signals whose
amplitude and phase vary as the earth rotates relative to the astronomical source. (Average daily variations of some important environmental parameters are given in Table~\ref{tab: daily var}). In addition to such ``false effects'', we evaluate scale factor uncertainties in $N_{\rm p}$, the number of polarized electrons in our pendulum.
%
%	Table I
%
\begin{table}[h]
\caption{Average daily components the systematic effects}
\begin{ruledtabular}
\begin{tabular}{cc}
effect & daily variation \\
\colrule
ambient magnetic field & $2.0\pm 0.4$ mG \\
tilt & $4.7\pm 2.1$ nrad \\
$Q_{22}$ gravity gradient & $(1.10\pm 0.27)\times 10^{-3}$ g/cm$^3$ \\
$Q_{21}$ gravity gradient & $(5.0\pm 3.4)\times 10^{-4}$ g/cm$^3$ \\
temperature drift & $0.34\pm 0.21$ mK \\
$1\omega$ temperature & $2.2\pm 1.7 ~\mu$K \\
\end{tabular}
\end{ruledtabular}
\label{tab: daily var}
\end{table}

\subsection{Turntable Rotation Rate}    

We observed an $\approx 80$ nrad  1$\omega$ signal that was very reproducible from cycle to cycle and independent of the orientation of the spin pendulum within the rotating apparatus. This arose primarily from a non-constant
rotation rate caused by imperfections 
in the turntable bearings and appeared as a twist signal 
because we measured the angular position of the ``inertial" pendulum relative to the rotating vacuum vessel.
Although a servo system locked the output of the turntable's rotary encoder 
to a crystal oscillator to produce a nominally constant rotation rate, the finite gain of the servo loop and
imperfections in the rotary encoder
allowed residual twist signals to appear at the harmonics of the turntable rotation frequency as illustrated in 
Fig.~\ref{fig: fft}. 

We model the true phase, $\phi$, of the turntable as
$$ \phi=\omega t + \sum_{n=1}^8 \phi_n e^{i \omega_n t} $$
where $\omega$ is the nominally constant angular frequency of the turntable, $\omega_n = n\omega$, and $\phi_n$ is the 
amplitude of the phase variation due to turntable imperfection. It is easy 
to show that $\phi_n$ will produce an
angular twist signal, $\theta_n$, at frequency $\omega_n$ (ignoring damping) given by
\begin{equation}
\theta_n = \phi_n \frac{\omega_n^2}{\omega_p^2 -\omega_n^2} 
\label{eq: ttsys}
\end{equation}
where $\omega_p$ is the free torsional angular frequency of the pendulum.

\subsubsection{Astronomical signals}

We searched for time variations, $\Delta \phi_n$, of the turntable imperfections whose amplitudes and phases project onto 
the time dependence of our astronomical sources by fitting the time sequence of the $\theta_4$ and $\theta_8$ values to 
the basis functions of each astronomical signal. We chose $\theta_4$ and $\theta_8$ 
because signals at 4$\omega$ and 8$\omega$ do not change as the pendulum is rotated to each of its four positions 
within the vacuum vessel.
We found no evidence
for a modulation of $\theta_4$ or $\theta_8$ at any of the astronomical frequencies. Furthermore, had there been a 
coherent modulation, by taking data at all four orientations of the spin pendulum within the vacuum vessel, the 
1$\omega$ signal from $\Delta \phi_1$ would have been reduced by a factor of $0.22$ (more data was taken on mirror 1 than on
the other mirrors). Assuming that $\Delta \phi_1 \approx \Delta \phi_4 \approx \Delta \phi_8$ as would be the case for a
stiff point on the turntable bearing, and using Eq.~\ref{eq: ttsys}
with the factor of $0.22$, the systematic error from $\Delta \phi_1$ is less than $0.5\%$ of the statistical error for
any of the astronomical sources. This error is significantly smaller than the other systematic errors in Tables~\ref{tab: syst error astro} and \ref{tab: syst error SME} and was dropped from further consideration.
%
%	Table II
%
%\begin{widetext}
\begin{table*}[!]
\caption{Systematic and random errors in the astronomically modulated $A$, $B$ and $C$ signals evaluated in the CMB rest frame. Corrections were only applied for magnetic effects; the remaining effects were assigned  systematic errors equal to the absolute value of the effects plus their 1$\sigma$ uncertainties. The total systematic errors are quadratic sums of the uncertainty in the magnetic correction and the systematics of the remaining effects. A 2.3\% scale-factor uncertainty is not yet included.}
\begin{ruledtabular}
\begin{tabular}{cccccccc}
\multicolumn{1}{c}{signal} & \multicolumn{6}{c}{systematic error [eV]} & \multicolumn{1}{c}{random error [eV]} \\
 & magnetic & tilt & grav. gradient  & temp. drift &  1$\omega$ temp. &  total  &  \\
\colrule
$A_X \times 10^{22}$     & $+0.20 \pm 0.07$  & $\pm 0.31$  & $ \pm 0.07$ &  $\pm 0.04$ &  $\pm 0.05$ &  $\pm 0.33$ & $\pm 1.31$\\
$A_Y \times 10^{22}$     & $-0.10 \pm 0.03$  & $\pm 0.24$  & $ \pm 0.20$ &  $\pm 0.03$ &  $\pm 0.02$ &  $\pm 0.32$ & $\pm 1.32$ \\
$B \times 10^{20}$       & $-1.40 \pm 0.46$  & $\pm 2.5$   & $ \pm 0.38$ &  $\pm 0.23$ &  $\pm 0.37$ &  $\pm 2.61$ & $\pm 11.0$ \\
$C_{XX}\times 10^{18}$   & $+0.95 \pm 0.31$  & $\pm 0.43$    & $ \pm 0.17$ &  $\pm 0.08$ &  $\pm 0.05$ &  $\pm 0.57$ & $\pm 2.16$ \\
$C_{XY}\times 10^{18}$   & $+2.29 \pm 0.76$  & $\pm 0.84$    & $ \pm 0.25$ &  $\pm 0.21$ &  $\pm 0.06$ &  $\pm 1.18$ & $\pm 4.24$ \\
$C_{XZ}\times 10^{18}$   & $-3.61 \pm 1.19$  & $\pm 2.58$   & $ \pm 0.99$ &  $\pm 0.31$ &  $\pm 0.22$ &  $\pm 3.03$ & $\pm 10.16$ \\
$C_{YX}\times 10^{18}$   & $-0.73 \pm 0.24$  & $\pm 0.44$    & $ \pm 0.33$ &  $\pm 0.04$ &  $\pm 0.06$ &  $\pm 0.60$ & $\pm 2.17$ \\
$C_{YY}\times 10^{18}$   & $-1.04 \pm 0.34$  & $\pm 0.76$    & $ \pm 0.67$ &  $\pm 0.15$ &  $\pm 0.07$ &  $\pm 1.08$ & $\pm 4.25$ \\
$C_{YZ}\times 10^{18}$   & $+4.02 \pm 1.33$  & $\pm 2.46$   & $ \pm 1.60$ &  $\pm 0.21$ &  $\pm 0.32$ &  $\pm 3.24$ & $ \pm 10.23$ 
\end{tabular}
\end{ruledtabular}
\label{tab: syst error astro}
\end{table*}
%\end{widetext}
%
%
\subsubsection{Lab-fixed signals}
We distinguished lab-fixed sources that couple to the spin dipole of the pendulum from 1$\omega$ signals that are
independent of the spin dipole by taking data at four equally spaced orientations
of the spin dipole within the apparatus. A systematic error can arise if $\phi_1$ in Eq.~\ref{eq: ttsys} changes as
the orientation of the spin dipole is rotated. We searched for such a dependence by comparing the $\theta_4$ and 
$\theta_8$ signals for the four orientations of the spin dipole and found no variation beyond the statistical error.
Assuming that $\Delta \phi_1 \approx \Delta \phi_4 \approx \Delta \phi_8$,
Eq.~\ref{eq: ttsys} predicts that $\Delta \phi_1$ produces a lab-fixed systematic 
of $0.11 \times 10^{-22}$ eV as listed 
in Table~\ref{tab: syst error lab-fixed}}.

%
%	Table III
%
%\begin{widetext}
\begin{table*}[!tbp]
\caption{Systematic and random errors in the heliocentrically modulated SME coefficients and one-boson exchange interactions. Corrections were only applied for magnetic effects; the remaining  effects were assigned systematic errors equal to the absolute value of effects plus their 1$\sigma$ uncertainty. The total errors are quadratic sums of the uncertainty in the magnetic corrections and the systematics of the other effects. A 2.3\% scale-factor uncertainty is not yet included.}
\begin{ruledtabular}
\begin{tabular}{cccccccc}
\multicolumn{1}{c}{signal} & \multicolumn{6}{c}{systematic error [eV]} & \multicolumn{1}{c}{random error [eV]} \\
 & magnetic & tilt & grav. gradient  & temp. drift &  1$\omega$ temp. &  total  &  \\

\colrule
$A_X \times 10^{22}$              & $-0.05\pm 0.02$   & $\pm 0.30$  & $\pm 0.06$ & $\pm 0.05$  & $\pm 0.05$ & $\pm 0.31$ & $\pm 1.44$ \\
$A_Y \times 10^{22}$              & $+0.08 \pm 0.03$  & $\pm 0.21$  & $\pm 0.17$  & $\pm 0.03$ & $\pm 0.02$ &  $\pm 0.27$& $\pm 1.44$ \\
$C_{XX}^{\prime}\times 10^{18}$   & $+0.94\pm 0.31$   & $\pm 0.44$  & $\pm 0.18$  & $\pm 0.08$ &  
$\pm 0.05$ &  $\pm 0.58$& $\pm 2.16$ \\
$C_{XY}^{\prime}\times 10^{18}$   & $+0.73\pm 0.24$   & $\pm 0.64$  & $\pm 0.23$  & $\pm 0.12$ &  
$\pm 0.06$ &  $\pm 0.73$& $\pm 1.98$ \\
$C_{YX}^{\prime}\times 10^{18}$   & $-0.73\pm 0.24$   & $\pm 0.44$  & $\pm 0.33$  & $\pm 0.04$ &  
$\pm 0.06$ &  $\pm 0.60$& $\pm 2.17$ \\
$C_{YY}^{\prime}\times 10^{18}$   & $+0.70\pm 0.23$   & $\pm 0.56$  & $\pm 0.08$  & $\pm 0.16$ &  
$\pm 0.11$ &  $\pm 0.64$& $\pm 1.99$ \\
${f_\perp}/(\hbar c)\times 10^{32}$               & $-0.15\pm 0.05$ & $\pm 0.04$ & $\pm 0.31$  & $\pm 0.11$ &  $\pm 0.04$  &  $\pm 0.34$& $\pm 2.09$ \\
${f_v}/(\hbar c)\times 10^{56}$                   & $-1.29\pm 0.43$ & $\pm 0.68$ & $\pm 0.14$  & $\pm 0.07$ &  $\pm 0.12$  &  $\pm 0.83$& $\pm 2.32$ \\
$g_{\rm P}^e g_{\rm S}^N/(\hbar c)\times 10^{37}$ & $-0.10\pm 0.03$ & $\pm 1.60$ & $\pm 0.40$  & $\pm 0.31$ &  $\pm 0.29$    &  $\pm 1.70$& $\pm 8.53$ 
\end{tabular}
\end{ruledtabular}
\label{tab: syst error SME}
\end{table*}
%\end{widetext}
%	the following blank line is important

%
%	Table IV
%
\begin{table}[h]
\caption{Error budget for lab-fixed signals.
% showing uncertainties in $|\beta|\times 10^{22}$. 
A scale-factor uncertainty equaling 18\% of the central value must be folded into the quadratic sum of the random and statistical errors.}
\begin{ruledtabular}
\begin{tabular}{lrrr}
 & \multicolumn{3}{c}{uncertainty in $\beta \times 10^{22}$~[eV]} \\
 date  & 10/2006  & 6/2007 & 3/2008 \\
%   & error [eV]	& error [eV] & error [eV]\\
\colrule
systematic effect & & \\
~~~tilt	& 0.52	& 0.88 & 0.36 \\
~~~gravity gradient	& 0.77	& 0.56 & 0.65\\
~~~temp drift	& 0.14	& 0.14 & 0.20 \\
~~~1$\omega$ temp  & 0.14  &  0.14 & 0.20 \\
~~~turntable speed	& 0.11	& 0.11 & 0.11 \\
~~~magnetic	& 1.06	& 1.06 & 0.51 \\
total systematic error	& 1.43	& 1.50 & 0.95 \\
\colrule
random error & 7.5	& 5.6  & 7.7\\
\end{tabular}
\end{ruledtabular}
\label{tab: syst error lab-fixed}
\end{table}
\subsection{Turntable Tilt}    
If the rotation axis of the turntable is tilted from local vertical, the upper fiber attachment point 
flexes as the apparatus rotates, leading to a ``tilt-twist" 1$\omega$ angular deflection of the pendulum. Imperfections in the mirrors of the system that measures our twist signal produce a related effect as the tilt causes the light spot to move on the mirrors. 
The tilt of our laboratory floor changes by a few $\mu$rad per day which, if not corrected, would produce a spurious 1$\omega$ signal 
that varies by up to $40$~nrad per day.
The
``feetback" control loop described in Appendix~\ref{ap: feetback} aligns the rotation axis to within $10$ nrad of the 
local vertical. 

We measure the tilt-twist feedthrough (the ratio of the spurious signal to the tilt) by applying known tilts to the apparatus; we find that the
feedthrough is typically $3\%$ with an amplitude and phase that depend upon the angular orientation of the 
pendulum within the apparatus. A conservative upper limit to the component of the tilt-twist feedthrough that 
mimics a coupling to the spin dipole
of the pendulum ({\it i.e.} that rotates along with the spin dipole) is $4\%$.

\subsubsection{Astronomical signals}

We extracted any tilt component that mimicked an astronomical signal by fitting the outputs from the rotating tilt sensors to the basis functions of each of the astronomical signals.  The largest projection, just
over 3 standard deviations, was found for the $C_{XY}$ basis function. We multiplied these tilt projections by
the tilt-twist feedthrough, $4\%$, to determine the systematic error due to tilt. The results, given in 
Table~\ref{tab: syst error astro}, show that control of the tilt-twist feedthrough is the largest source of systematic error for the 
astronomical signals.

\subsubsection{Lab-fixed signals}

The average tilt of the apparatus was 2.2, 3.7 and 1.5 nrad for lab-fixed Measurements 1, 2 and 3, respectively.
The component of the tilt-twist feedthrough that rotated with the orientation of the spin dipole was no more than
$2\%$ for these data sets. The lab-fixed systematic error from tilt is then $2\%$ of measured tilt, given in
Table~\ref{tab: syst error lab-fixed}.

\medskip

\subsection{Magnetic Coupling}    
\medskip

The spin pendulum had a residual magnetic dipole moment of $7 \times 10^{-3}$ erg/G and was located within four
layers of magnetic shielding. When the currents to the Helmholtz coils outside of the apparatus were reversed, a 
horizontal field of $374$ mG was applied to the apparatus which produced a 
1$\omega$ signal with a typical magnitude of $20 \pm 3$ nrad ($\beta=2 \times 10^{-20}$ eV). 
The spin pendulum was insensitive to reversal of the current in the vertical Helmholtz coil with
a feedthrough of less than $2 \pm 3$ nrad for a $470$ mG change in vertical field.

\subsubsection{Astronomical signals}

\begin{figure}[h]
\hfil\scalebox{0.47}{\includegraphics*[53pt,31pt][559pt,377pt]{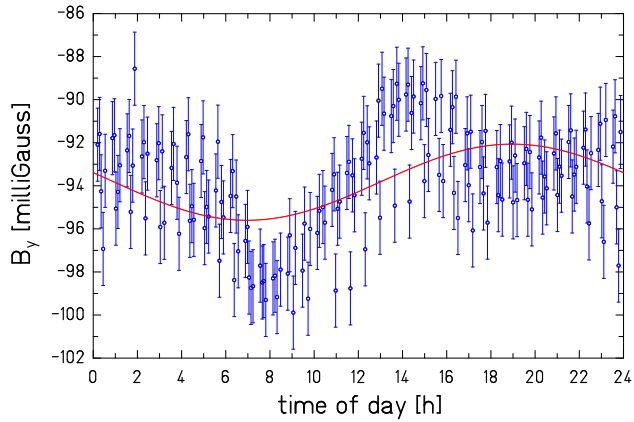}}\hfil
\caption{Daily variation of the ambient magnetic field measured with a flux-gate magnetometer.} 
\label{fig: daily mag}
\end{figure}

Before taking the data presented in this paper, as well as during the last $20\%$ of the data, a 
triple-axis fluxgate magnetometer
was placed within the Helmholtz coils (but outside of the heat shield) to monitor the stability of the 
magnetic environment. We were surprised to find an average daily variation of the horizontal magnetic field 
of $2.0 \pm 0.4$ mG (see Fig.~\ref{fig: daily mag}).
Because the astronomical signals have a large daily component, the magnetic coupling to daily field variations could lead to
systematic errors as large as $56\%$ of the statistical error. To reduce this error, we corrected our data for the coupling
to the daily variation of the magnetic field. No other corrections to the 
astronomical signals were required. 

%
%	Figure 12
%
The amplitude and phase of the magnetic coupling
to the spin pendulum were found by 
reversing the Helmholtz coil currents, while the fluxgate measurements provided us with the amplitude and phase of the daily variation of the
horizontal magnetic field. The calculated magnetic coupling was subtracted from each data cut before fitting the data to
the astronomical signal basis functions. We assigned a systematic error for this procedure as follows. 
For each astronomical signal, we fitted the 
data with and without the magnetic correction to determine the magnitude of the correction (always less than $56\%$ of the 
statistical error). There was a $15\%$ uncertainty in the magnetic feedthrough calibration and a
$29\%$ uncertainty in the amplitude of the daily magnetic field variation 
(obtained from the difference between the measurements 
before and at the end of the data sequence). We added these uncertainties in quadrature to obtain magnetic systematic errors equal to 33\% of the magnetic corrections. Results  are listed in Tables \ref{tab: syst error astro} and \ref{tab: syst error SME}. 
\subsubsection{Lab-fixed signals}
An applied horizontal
field of $374$ mG gave a magnetic 1$\omega$ signal, $S_M$, of $22.2 \pm 1.4$ nrad in Measurements 1 and 2, and $10.8\pm2.4$ nrad in Measurement 3. The Helmholtz coil currents were adjusted to cancel the local field to within $0.5$ mG with an 
uncertainty of $1$ mG due to the daily variation of the horizontal field. We compute the magnetic lab-fixed systematic error
from $1.5/374 \times S_M$, {\em i.e.} 0.089 nrad ($\beta = 1.06 \times 10^{-22}$ eV) in Measurements 1 and 2 and 0.043 nrad in Measurement 3.

\subsection{Gravity Gradients}    

We use a spherical multipole basis\cite{su:94} to characterize the mass moments, $q_{lm}$, of the spin pendulum. The coupling of
$q_{l1}$ mass moments to $Q_{l1}$ gravity gradient fields produce 1$\omega$ angular deflections of the pendulum that
have the same dependence upon the orientation of the pendulum as would a coupling to the spin dipole. Although
the spin pendulum was designed to have vanishing $q_{l1}$ moments, imperfections produced small residual moments. The
only significant gravity gradient coupling for this work was $q_{21} Q_{21}$. This coupling was minimized by using
a special gradiometer pendulum with a large $q_{21}$ moment to measure ambient $Q_{21}$ field. We mounted machined 
Pb/brass masses on a turntable just outside of the vacuum vessel to cancel the local $Q_{21}$ field by a 
factor of $200$. Finally, we temporarily rotated the Pb/brass masses to double the gradient,  measured the residual $q_{21}$ moment of the 
pendulum, and adjusted tuning screws on the pendulum (see Fig.~\ref{fig: spin pendulum}) to minimize the 
residual $q_{21}$ mass moment. By canceling the $Q_{21}$ field and minimizing the $q_{21}$ moment of the pendulum, the 
$q_{21} Q_{21}$ coupling was reduced to tolerable levels.

\subsubsection{Astronomical signals}

We were concerned that daily variations of the local $Q_{21}$ field
might couple to the residual $q_{21}$ moment of the pendulum to produce a daily modulated 
1$\omega$ signal that would project onto the astronomical signal basis functions. The $Q_{22}$ field from local sources 
is typically larger than the $Q_{21}$ field (because $Q_{22}$ is maximum at a polar angle of $90^{\circ}$ where $Q_{21}$ vanishes) and 
falls off with distance with the same radial dependence as the $Q_{21}$ field. The pendulum had a residual $q_{22}$ moment (uncompensated) that
was $7$ times larger than the residual $q_{21}$ moment. We used the $q_{22} Q_{22}$ coupling to monitor the time 
variation of the $Q_{22}$ field which for local sources will be correlated with changes in the $Q_{21}$ field.
The $q_{22} Q_{22}$ coupling produces a 2$\omega$ signal.
We fitted the time sequence of the measured 2$\omega$ signals to the astronomical
basis functions to determine the projection of any $Q_{22}$ variations onto those functions. We then ran a correlation 
analysis between the 1$\omega$ and 2$\omega$ signals over the entire data set to extract the linear slope between the
1$\omega$ and 2$\omega$ signals. That slope was found to be less than $3\%$. The gravity gradient systematic error
in Table~\ref{tab: syst error astro} is the projection of the 2$\omega$ signal onto the basis functions times $0.03$.

An independent analysis of the daily variation of the $Q_{21}$ field was made using $8$ day-long data runs with the 
gradiometer pendulum. The gradiometer pendulum had a $q_{21}$ moment $90$ times larger than the residual moment of the
spin pendulum. No evidence for a daily variation in the $Q_{21}$ field was found with an uncertainty $4$ times larger that
obtained from the $q_{22} Q_{22}$ analysis.

\subsubsection{Lab-fixed signals}

When the Pb/brass $Q_{21}$ compensation masses were rotated to add to the local $Q_{21}$ gradient, a change in 
1$\omega$ signal of $70$ nrad typically was measured. This coupling would change each time the ball-cone attachment on
the spin pendulum was rotated because of small changes in the tilt of the pendulum relative to the ball-cone.
The residual 1$\omega$ signal when the Pb/brass compensators
canceled the local gradient was therefore less than $0.2$ nrad, typically $20\%$ of the statistical error for each 
ball-cone configuration. Because we knew the amplitude and phase of the total $Q_{21}$ field and the amplitude and
phase of the residual $q_{21}$ moment for each ball-cone configuration, we corrected the raw data for the residual
$q_{21} Q_{21}$ coupling. This was the only correction applied to the lab-fixed signals and never exceeded $30\%$ of the 
statistical error for any configuration. The largest uncertainty in this procedure was the assumption that the total
$Q_{21}$ field at the pendulum did not change appreciably with time. We assign a systematic error to this procedure by
taking the difference between the lab-fixed signals with and without the $q_{21} Q_{21}$ correction applied, and 
multiplying this difference by $50\%$ to account for possible changes in the $Q_{21}$ field. 
\subsection{Thermal Effects}    
The equilibrium twist angle of the spin pendulum is a strong function of temperature. 
Because these and other thermal effects are difficult to model, we took special care to stabilize and monitor the thermal environment of the apparatus.
Four temperature sensors were mounted on the 
rotating apparatus, four others were attached to stationary components of the apparatus, and two sensors monitored the lab air temperature. Constant-temperature water from one Neslab RTE-221 unit flowed through copper pipes soldered to copper
heat shields that surround the apparatus, and water from a second unit flowed through radiators with fans to hold constant the 
temperature of the room that houses the apparatus. The resulting thermal environment for the pendulum is constant 
to within a few mK per day.

We consider two thermal effects. The first is the direct effect of a temperature change on the 1$\omega$
signal of the pendulum. We measured that a $1$ K change in temperature of the apparatus caused a $9 \pm 4$ nrad change in the
1$\omega$ signal ($\beta =1.7 \times 10^{-20}$ eV/K averaged over the sensors). 
Second, the temperature sensors on the rotating apparatus show small variations ($\approx 0.1$ mK) at
the turntable rotation frequency. Part of this signal may be due to friction in the turntable bearings which, if changing
in time, could lead to changes in the 1$\omega$ signal of the pendulum.

\subsubsection{Astronomical signals}

We determined the direct temperature effect by fitting the measured mean temperatures of the sensors on the apparatus to the astronomical
basis functions. We then multiplied these by the measured
feedthrough, $1.7 \times 10^{-20}$ eV/K, to obtain the systematic errors given as ``temp. drift"
in Table~\ref{tab: syst error astro}.
Similarly, we found the 1$\omega$  temperature effect by fitting the 1$\omega$ signals of the rotating temperature sensors to the 
astronomical basis functions and multiplied by the feedthrough. 
We determined the feedthrough by performing a correlation analysis between the 1$\omega$
signals on the rotating temperature sensors and the 1$\omega$ signal of the pendulum. The feedthrough was found to be 
$\le 160  \times 10^{-20}$ eV/K. The systematic error is given as ``$1\omega$ temp." in Table~\ref{tab: syst error astro}.

\subsubsection{Lab-fixed signals}

The standard deviation of the average temperature of the sensors on the apparatus for the lab-fixed data runs was
$2.8$ mK. Because there was no correlation between the average temperature and the orientation of the spin pendulum
within the apparatus, the direct temperature feedthrough created a random variation of the 1$\omega$ pendulum
signal. The systematic error for the temperature drift effect, given in Table~\ref{tab: syst error lab-fixed} is $2.8\ {\rm mK} \times 1.7 \times10^{-20}\ 
{\rm eV/K} /\sqrt{N-1}$ where $N$ is the number of data runs. In measurements 1 and 2, $N=16$, while measurement 3 had $N=8$.
The standard deviation of 1$\omega$ signals on the rotating temperature sensors for the lab fixed data runs was
$28\ \mu$K. Again, these signals were not correlated with the orientation of the spin pendulum and would produce a 
random error. We used the same feedthrough for the bearing friction temperature effect as above for the astronomical 
signals to assign a systematic error for the $1\omega$ temperature feedthrough of $28\ \mu {\rm K} \times 160 \times 10^{-20}\ {\rm eV/K} /\sqrt{N-1}$ for the
lab-fixed measurements.

\subsection{Spin content of the pendulum}
%Calibration of $\bm N_{\rm p}$}
\label{sec: Np calib}
The uncertainty in the pendulum's spin content adds a scale-factor uncertainty to all our constraints.
\subsubsection{Lab-fixed signals}
Our constraints on $A_Z$, $C_{ZX}$, $C_{ZY}$, $C_{ZZ}$ and 
the $CP$-violating interaction of Eq.~\ref{eq: MW} with $\lambda << 1\,$A.U. are derived from laboratory-fixed signals. In this case we adopt the spin content and its 18\% uncertainty given in Eq.~\ref{eq: Npol}.

\subsubsection{Astronomical signals}
Our constraints on these signals are based on effects occuring  at sidereal or solar frequencies, and steady lab-fixed ${\bm \beta}$'s were ignored in the analysis. This allowed us to evaluate the constraints using a more precise experimental value for the spin content based on the fact that
it is extremely unlikely that new fundamental physics could have produced the lab-fixed signal in Table~\ref{tab: Phase II lab fixed}. This would require either a preferred-frame $\bm A$ that happened to point exactly opposite to the Earth's spin axis (this has  
a probability 
$\Delta \Omega/(4\pi)\approx[\delta \beta_{\rm E}\cos \Psi /(2 \beta_{\rm N})]^2=7\times10^{-5}$) or a $CP$-violating interaction of Eq.~\ref{eq: MW} with 
$R_{\oplus} \le \lambda \le 3\!\times \! 10^6$ km. If we neglect these unlikely possibilities, our measured torque $\kappa\delta_{\rm N}=-(0.2537\pm 0.0057)$ fN$\,$m provides, via Eq.~\ref{eq: gyro torque}, an absolute calibration of the spin content of our pendulum,
\begin{equation}
\label{eq: Np from gyro}
N_{\rm p}^{\rm exp}=(9.80 \pm 0.22)\times 10^{22}~,
\end{equation}
which gives a 2.3\% scale-factor uncertainty for the astronomically signals. This was always negligible incomparison to the statistical error.
\section{Results}
\label{sec: results}
In this section all constraints given in numerical form are $1\sigma$ and include both random and systematic errors. Constraints presented in graphical form are at the 95\% confidence level. 
\subsection{Preferred-frame constraints}
\subsubsection{Phenomenological cosmic preferred frames}
To constrain the parameters ${\bm A}$ and $\bm{C}$ we require results from lab-fixed as well as astronomically modulated signals because the $A_Z$, $C_{ZX}$, $C_{ZY}$ and $C_{ZZ}$ signatures have no sidereal modulation. The lab fixed results are displayed in Table~\ref{tab: Phase II lab fixed}; the combined result of the three measurements is
\begin{eqnarray}
\hat{\beta}_{\rm N}&=&(-0.03\pm 0.28)\times 10^{-20}~{\rm eV} \nonumber \\
\beta_{\rm E}&=&(0.00\pm 0.04)\times 10^{-20}~{\rm eV}~, 
\label{eq: comb lab-fixed}
\end{eqnarray}
where $\hat{\beta}_{\rm N}\equiv\beta_{\rm N}-\beta_{\rm N}^{\rm gyro}$, and the overall scale factor error from the uncertainty in $N_{\rm p}$ is now included.
The results in Table~\ref{tab: Phase II lab fixed} allow us to extract values for either $A_Z$, or else for $C_{ZX}$, $C_{ZY}$ and $C_{ZZ}$.
The result in Eq.~\ref{eq: comb lab-fixed} implies that
\begin{eqnarray}
\label{eq: A_z extraction}
A_{\rm Z}&=&\hat{\beta}_{\rm N}/\cos\Psi  \nonumber \\
 &=& (-0.4\pm 4.4)\times 10^{-21}~{\rm eV}~.
\end{eqnarray}
On the other hand, to constrain the boost terms we must invert the equations
\begin{eqnarray}
\tilde{v}_X(t_1) \;C_{ZX}\!+\!\tilde{v}_Y(t_1)\;\tilde{C}_{ZY}\!+\!\tilde{v}_Z(t_1)\;\tilde{C}_{ZZ}\!&=&\! \hat{\beta}_{\rm N}(t_1)/\cos\Psi  \nonumber \\
\tilde{v}_X(t_2) \;C_{ZX}\!+\!\tilde{v}_Y(t_2)\;\tilde{C}_{ZY}\!+\!\tilde{v}_Z(t_2)\;\tilde{C}_{ZZ}\!&=&\!\hat{\beta}_{\rm N}(t_2)/\cos\Psi \nonumber \\
\tilde{v}_X(t_3) \;C_{ZX}\!+\!\tilde{v}_Y(t_3)\;\tilde{C}_{ZY}\!+\!\tilde{v}_Z(t_3)\;\tilde{C}_{ZZ}\!&=&\!\hat{\beta}_{\rm N}(t_3)/\cos\Psi~. \nonumber
\label{eq: C_0 def}
\end{eqnarray}
where $\tilde{v}$ is the laboratory velocity in units of $c$ and $t_i$ refers to the mean time of a lab-fixed measurement given in 
Table~\ref{tab: Phase II lab fixed}.
We obtain
\begin{eqnarray}
C_{ZX} &=& (-4.9 \pm 8.8 \pm 0.9) \times 10^{-18}~{\rm eV} \nonumber \\
C_{ZY} &=& (-9.6 \pm 18.9 \pm 8.8) \times 10^{-18}~{\rm eV} \nonumber \\
C_{ZZ}&=& (+24.8 \pm 38.6 \pm 20.9) \times 10^{-18}~{\rm eV} ~,
\label{eq: A_z & C_0 values}
\end{eqnarray}
where the second error reflects the scale factor uncertainty.
%
%	Table V
%
\begin{table}[t]
\caption{Three separate measurements of lab-fixed signals. Units of torque, $\kappa \,\delta$ and $\beta=\kappa\delta/N_{\rm p}$ are 
 $10^{-16}$ N\,m and $10^{-20}$~eV, respectively. Dates refer to the mean time of the measurements.
Errors in $\beta$ include the systematic and random errors given in Table~\ref{tab: syst error lab-fixed} but not the scale-factor uncertainty.} 
%xxx program del2beta xxx}
\begin{ruledtabular}
\begin{tabular}{llll}
signal  & 15/10/06 value  & 15/6/07 value & 8/3/08 value\\
\colrule
$\kappa \, \delta_{\rm N}$  & $-2.49 \pm 0.11$ &  $-2.57 \pm 0.08$  &  $-2.52\pm 0.12$ \\
$\beta_{\rm N}$ &   $-1.62 \pm 0.07$  &  $-1.67\pm 0.05$  &  $-1.64 \pm 0.08$\\
$\beta_{\rm}^{\rm gyro}$ & $-1.62$ & $-1.62$ & $-1.62$\\
$\beta_{\rm N}-\beta_{\rm}^{\rm gyro}$ & $+0.00 \pm 0.07$ & $-0.05 \pm 0.05$ & $-0.02 \pm 0.08$\\
\colrule
$\kappa \, \delta_{\rm E}$  & $+0.04 \pm 0.11$ &  $-0.06 \pm 0.08$  &  $+0.08 \pm 0.12$ \\
$\beta_{\rm E}$ &   $+0.03 \pm 0.07$   &  $-0.04\pm 0.05$  & $+0.05 \pm 0.08$\\
\end{tabular}
\end{ruledtabular}
\label{tab: Phase II lab fixed}
\end{table}

Our complete constraint on the electron's Lorentz-symmetry violating rotation parameter $\bm{A}$ is displayed in Table \ref{tab: A params};
it is roughly two orders of magnitude more restrictive than previous work by Hou {\em et al.}\cite{ho:03}, 
%
%	Table VI
%
\begin{table}[!h]
\caption{$1\sigma$ constraints from our work and from Hou et al.\cite{ho:03} on the Lorentz-violating rotation parameters defined in Eq.~\ref{eq: def A}.
Units are $10^{-22}$ eV. The errors in $A_Z$ are larger than those in $A_X$ and $A_Y$ because of the greater systematic uncertainty in lab-fixed signals. It is assumed that $C$ terms can be neglected.}
\begin{ruledtabular}
\begin{tabular}{ccc}
parameter  & this work    &  Hou et al.\cite{ho:03}  \\
\colrule
 \\
$A_X$  &   $-0.67\pm 1.31$  &   $-108\pm 112$ \\
$A_Y$  &   $-0.18\pm 1.32$  &   $-5\pm 156$ \\
$A_Z$  &   $-4\pm 44$     &   $107\pm 2610$ \\
\end{tabular}
\end{ruledtabular}
\label{tab: A params}
\end{table}
%
%
%	Table VII
%
\begin{table}[h!]
\caption{$1\sigma$ constraints on the $C$ parameters of Eq.~\ref{eq: def C}. The last three rows in this table are inferred from lab-fixed measurements. Units are $10^{-18}$ eV and the scale factor uncertainty is included.}
\begin{ruledtabular}
\begin{tabular}{lrlrlr}
parameter & value  & parameter & value \\
\colrule
\\
$C_{XX}$  & $\!\!+0.96 \pm 2.16$  &   $C_{YX}$  & $\!\!-3.74\pm 2.17$    \\
$C_{XY}$  & $\!\!+1.84\pm 4.24$   &   $C_{YY}$  & $\!\!-7.76\pm 4.25$    \\
$C_{XZ}$  & $\!\!-3.92\pm 10.16$  &   $C_{YZ}$  & $\!\!+16.17\pm10.23$   \\
$C_{ZX}$  & $\!\!-4.92 \pm 8.86$    &   & & \\
$C_{ZY}$  & $\!\!-9.6\pm 20.9$ \\
$C_{ZZ}$  & $\!\!+24.8 \pm 43.9$    &  & 
\end{tabular}
\end{ruledtabular}
\label{tab: C params}
\end{table}
and substantially below the benchmark value 
$m_e^2/M_{\rm Planck}= 2 \times 10^{-17}$~eV. 

Our $1\sigma$ constraint on the electron's helicity-generating parameter $B$ is
%defined in Eq.~\ref{eq: def B} 
%
\begin{equation}
B=(+0.50 \pm 1.13)\times 10^{-19}\;\rm{eV}~. 
\label{eq: B value}
\end{equation}
Table~\ref{tab: C params} displays our limits on the electron's generalized helicity tensor $\bm{C}$. 
We are not aware of any comparable measurements of either $B$ or $C$.

\subsubsection{Standard Model Extension}
The minimal Standard Model Extension or SME incorporates the possibility of Lorentz and $CPT$ violation by invoking pervasive, feeble, static spin 1 and spin 2 fields that are defined in the heliocentric frame (for electrons in Minkowski spacetime 44 free parameters are involved). 
 
We follow conventional experimental practice\cite{ho:03,ph:01,hu:03,ca:04} and first ignore possible violations of boost symmetry.
Then our constraints on ${\bm A}$ (Table~\ref{tab: A params}) translate directly to constraints on the SME parameters $\bm{\tilde b}^e$. 
Table \ref{tab: Kost params e,p,n} compares the electron's $\bm{\tilde b}$ parameters to the corresponding  proton and neutron parameters determined from hydrogen maser\cite{ph:01,hu:03} and dual $^{129}$Xe/$^3$He maser measurements\cite{ca:04}. The neutron and electron parameters have comparable upper limits. The hydrogen maser results have an uncertainty $10^4$ times greater so, in effect, they place a substantially weaker limit on $\bm{\tilde b}$ of the proton.
%
%	Table VIII
%
\begin{table}[h]
\caption{$1\sigma$ constraints on the Lorentz-violating rotation parameters for electrons, protons and neutrons. The SME boost terms are assumed to be negligible.
Units are $10^{-22}$ eV. Proton and neutron results are taken from Refs.~\cite{ph:01},\cite{hu:03} and Ref. ~\cite{ca:04}, respectively. }
\begin{ruledtabular}
\begin{tabular}{cccc}
parameter  & electron    &  proton & neutron \\
\colrule
 \\
$\tilde{b}_X$  &   $-0.67\pm 1.31$   &  $\leq 2\times 10^{4}$  & $0.22\pm 0.79$\\
$\tilde{b}_Y$  &   $-0.18\pm 1.32$   &  $\leq 2\times 10^{4}$  & $0.80\pm 0.95$ \\
$\tilde{b}_Z$  &   $-4\pm 44$    &   &  \\
\end{tabular}
\end{ruledtabular}
\label{tab: Kost params e,p,n}
\end{table}

We constrained boost terms in the SME by reevaluating our limits after 
setting the velocity of the sun to zero. The resulting boost-dependent coefficients are denoted by primes. 
%probe linear combinations of the SME parameters. 
The sidereally modulated data were fitted with 6 free parameters, $A_X$, $A_Y$, $C_{XX}^{\prime}$, $C_{XY}^{\prime}$, $C_{YX}^{\prime}$ and $C_{YY}^{\prime}$. 
In addition, we used the lab-fixed results in Table~\ref{tab: Phase II lab fixed} to constrain $A_Z$, $C_{\rm ZX}^{\prime}$ and 
\begin{equation}
\bar{C}_{ZY}^{\prime}\equiv C_{ZY}^{\prime}+\tan\epsilon\, C_{ZZ}^{\prime}~.
\end{equation} 
The $A$ and $C^{\prime}$ coefficients determine\cite{ak:03} the following combinations of SME parameters
\begin{eqnarray}
A_X&\!=&\tilde{b}_X^e~~~~~~~\;\;\;\;\;\;\;\;\;\;\;\;\;\;\;~~~~
A_Y\!=\tilde{b}_Y^e \nonumber \\
C^{\prime}_{XX}&\!=& [\tilde{b}_T+\tilde{d}_{-}-2\tilde{g}_c-3\tilde{g}_T+4\tilde{d}_{+}-\tilde{d}_Q]/2 \nonumber\\
C^{\prime}_{XY}&\!=& (\tilde{d}_{XY}-\tilde{H}_{ZT})+\tan{\epsilon}\,\tilde{H}_{YT} \nonumber\\
%C^{\prime}_{xz}&\!=& \sin{\epsilon}\tilde{H}_{YT} \nonumber \\
C^{\prime}_{YX}&\!=& \tilde{H}_{ZT}\nonumber\\
C^{\prime}_{YY}&\!=& \![2\tilde{g}_c-\tilde{g}_T-\tilde{b}_T+4\tilde{d}_{+}-\tilde{d}_- -\tilde{d}_Q]/2+ \nonumber \\
&~&~\tan\epsilon\,[\tilde{d}_{YZ}-\tilde{H}_{XT}] \nonumber \\
%C^{\prime}_{yz}&\!=&\sin\epsilon[\tilde{d}_{YZ}-\tilde{H}_{XT}]  \nonumber \\
C^{\prime}_{ZX}&\!=& \tilde{H}_{YT}-\tilde{d}_{ZX}~~\;\;\;\;\;\;\;\;\;\;~~
C^{\prime}_{ZY}\!= -\tilde{H}_{XT}  \nonumber\\
C^{\prime}_{ZZ}&\!=& [\tilde{g}_T-2\tilde{d}_+ +\tilde{d}_Q]~.
\end{eqnarray}
Our values for these parameters are listed in Table~\ref{tab: B hat e,n}.
%
%Table IX
%
\begin{table}[!h]
\caption{$1\sigma$ constraints on electron SME parameters from this experiment. Scale factor errors are included. The $A$ coefficients listed here differ from those in Table~\ref{tab: A params} because here we included additional boost terms whose functional forms are not fully orthogonal to those of the $A$ coefficients.}
\begin{ruledtabular}  
\begin{tabular}{lr}
parameter  & electron value (GeV) \\
\colrule
 \\
$A_X$  &   $(-0.9\pm1.4)\times 10^{-31}$   \\
$A_Y$  &   $(+0.9\pm1.4)\times 10^{-31}$   \\
$A_Z$  &   $(-0.3\pm 4.4)\times 10^{-30}$      \\
$C^{\prime}_{XX}$  &   $(+0.9\pm2.2)\times 10^{-27}$  \\
$C^{\prime}_{XY}$  &   $(+0.1\pm1.8)\times 10^{-27}$  \\
$C^{\prime}_{YX}$  &   $(-4.1\pm2.4)\times 10^{-27}$  \\
$C^{\prime}_{YY}$  &   $(-0.8\pm2.0)\times 10^{-27}$  \\
$C^{\prime}_{ZX}$  &   $(-4.9\pm 8.9)\times 10^{-27}$    \\
$\bar{C}^{\prime}_{ZY}$  &   $(+1.1\pm 9.2)\times 10^{-27}$    \\
\end{tabular}
\end{ruledtabular}
\label{tab: B hat e,n}
\end{table}
\subsubsection{Ghost condensate gravity}
\begin{figure}[h]
\hfil\scalebox{0.45}{\includegraphics*[55pt,39pt][568pt,442pt]{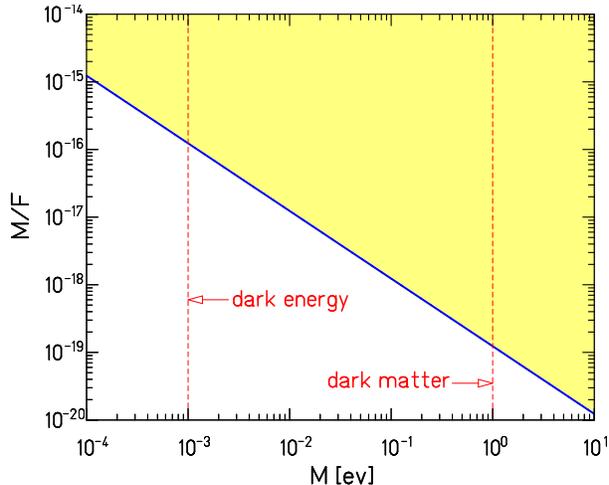}}\hfil
\caption{(color online) Upper limits on the ``ghost condensate'' parameter $M/F$ as a function of $M$. The shaded area is excluded at 95\% confidence. The vertical dashed lines show values of $M$ having particular cosmological significance.} 
\label{fig: Thaler}
\end{figure}

The result in Eq.~\ref{eq: B value} provides a limit on $-M^2/F$ (see Eq.~\ref{eq: Thaler mu}) of
\begin{equation}
\frac{M^2}{F} = (-0.50 \pm 1.10)\times 10^{-19}~{\rm eV}~,
\end{equation}
which improves on the value given in Ref.~\cite{ch:06} by three orders of magnitude. The corresponding 95\% confidence constraint on the ``ghost condensate parameter'' $M/F$ is shown as a function of $M$ 
in Fig.~\ref{fig: Thaler}. A condensate-mediated spin-spin interaction with a strength comparable to gravity would have the ghost condensate parameter $M/F\sim 10^{-16}$\cite{ch:06}.
%
%	Figure 13
%
\subsubsection{Non-commutative geometry}
Inserting the constraint on $|\bm{A}|$ from Table~\ref{tab: A params} into Eq.~\ref{eq: non-com pot}, we obtain an $1\sigma$ upper limit on $|\Theta|$, the minimum observable patch of area,
\begin{eqnarray}
|\Theta_{XZ}| ~{\rm or}~ |\Theta_{YZ}| \leq  4.9 \times 10^{-59} \;{\rm m^2}\left[{\rm 1~TeV}/\Lambda\right]^2 \\
|\Theta_{XY}| \ \leq  1.5 \times 10^{-57} \;{\rm m^2}\left[{\rm 1~TeV}/\Lambda\right]^2
\end{eqnarray}
which, assuming $\Lambda=1$~TeV, corresponds to energy scales of $2.8 \times 10^{13}$ GeV for $\Theta_{XZ}$ or 
$\Theta_{YZ}$ and $5.0 \times 10^{12}$ GeV  for $\Theta_{XY}$.
\subsubsection{Axial torsion}
Many theorists (see Refs.~\cite{sh:02} for a recent review) have speculated that the Reimann geometry of general relativity should be replaced by the more general Reimann-Cartan geometry that contains an additional non-symmetric field called torsion. Torsion does not directly couple to unpolarized matter and hence is difficult to detect; however torsion's fully-antisymmetric axial component 
$\bm {{\cal A}}$ is minimally
coupled through covariant derivatives to fermion spins. L\"ammerzahl\cite{la:97} noted that results from Hughes-Drever experiments can place useful bounds on $\bm{{\cal A}}$, and 
Kosteleck\'y, Russell and Tasson\cite{ko:07} recently showed how experiments, such as the one we report here, determine all three spatial components of $\bm {{\cal A}}$. Following Ref.~\cite{ko:07}, we find for minimal coupling (Ref.~\cite{ko:07} considers more complex cases as well) that our results in Table \ref{tab: A params} constrain any background torsion field (assumed to be roughly constant over the region of the solar system) to 
\begin{eqnarray}
{\cal A}_X&=& -4 A_X/(3\hbar c) = (+4.5 \pm 8.9)\times 10^{-16}~{\rm m}^{-1}~~\;\;\;\\
{\cal A}_Y&=& -4 A_Y/(3\hbar c) = (+1.2 \pm 8.9)\times 10^{-16}~{\rm m}^{-1}\nonumber \\
{\cal A}_Z&=& -4 A_Z/(3\hbar c) = (+2.7 \pm 29.8)\times 10^{-15}~{\rm m}^{-1}~.\nonumber
\end{eqnarray}
\subsection{Boson-exchange constraints}
\subsubsection{$CP$-violating monopole-dipole interactions}
\begin{figure}[h!]
\hfil\scalebox{0.48}{\includegraphics*[59pt,33pt][568pt,425pt]{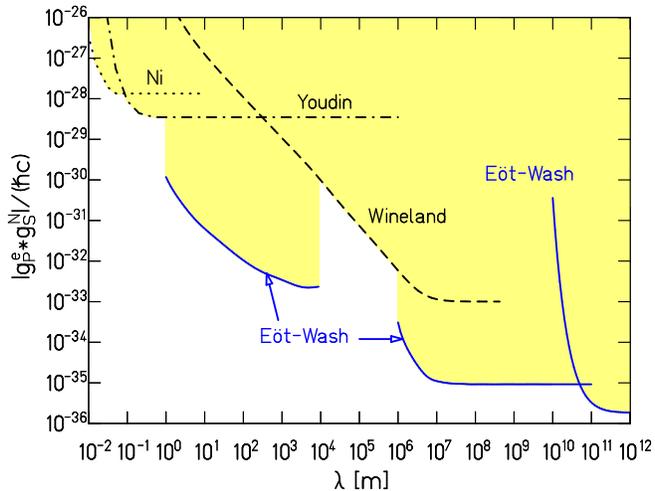}}\hfil
\caption{(color online) Upper limits on $|g_{\rm P}^e g_{\rm S}^N|/(\hbar c)$ as a function of interaction range $\lambda$; the shaded region is excluded at 95\% confidence. Our results and previous work by Youdin et al.\cite{yo:96}, Ni et al.\cite{ni:99} and Wineland et al.\cite{wi:91} are indicated by solid, dash-dotted, dotted and dashed lines, respectively.  
Our work does not provide constraints for 10 km $< \lambda < 10^3$ km because integration over the terrestrial surrounding is not reliable in this regime (see Ref.~\cite{ad:90}).} 
\label{fig: constraints}
\end{figure}

The constraints on $CP$-violating monopole-dipole interactions extracted from our lab-fixed and solar and lunar-source data are shown in Fig.~\ref{fig: constraints} (see also Table~\ref{tab: Dobr params}).
This new work improves over previous work\cite{yo:96,ni:99,wi:91} by factors of up to $10^4$.
\subsubsection{Velocity-dependent interactions}
\begin{figure}[!]
\hfil\scalebox{0.78}{\includegraphics*[47pt,45pt][328pt,428pt]{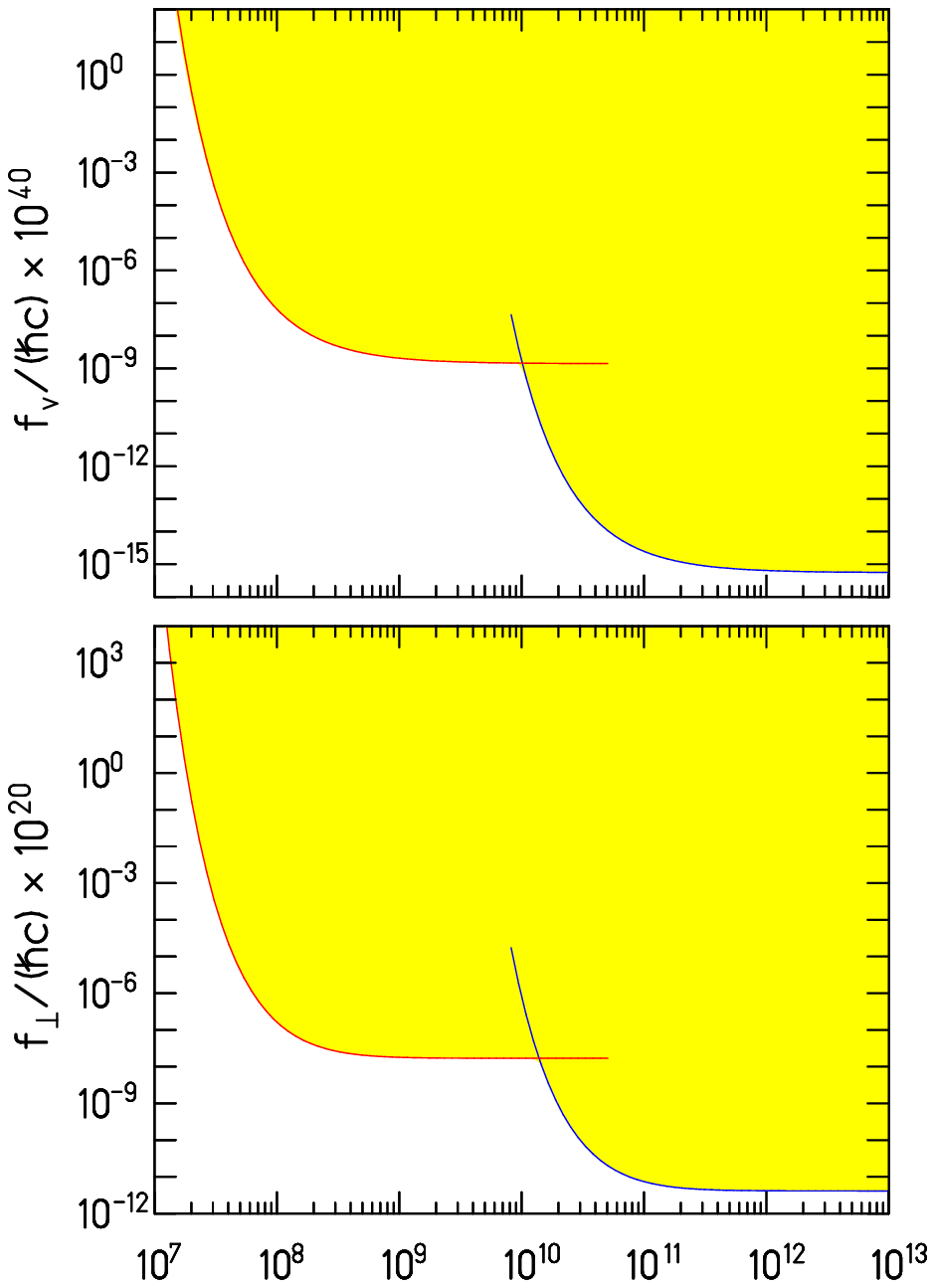}}\hfil
\caption{(color online) Constraints on the velocity-dependent couplings $f_v$ and $f_{\perp}$ inferred from the infinite-range results in Table~\ref{tab: Dobr params}. The shaded area is excluded at 95\% confidence. In making this plot from the $\lambda=\infty$ results in Table~\ref{tab: Dobr params} we neglected the eccentricities of the earth's and moon's orbits.} 
\label{fig: Fermilab range dependence}
\end{figure}

The limits on velocity-dependent boson-exchange forces, extracted from our experimental bounds on interactions between the spin pendulum and the sun are given in Table~\ref{tab: Dobr params}.
Figure \ref{fig: Fermilab range dependence} shows range-dependence of the velocity-dependent couplings $f_v$ and $f_{\perp}$. We are not aware of any previous measurements of such forces.

%
%	Table X
%
\begin{table}[h]
\caption{$1\sigma$ boson-exchange constraints from interactions with the Sun and Moon. Note that $f_v=2g_{\rm A}^e g_{\rm V}^N$. The solar and lunar constraints assume $\lambda >> 1.5\times 10^{11}$ m and $\lambda >> 4\times 10^{8}$ m, respectively.}
\begin{ruledtabular}
\begin{tabular}{lcc}
parameter  & solar constraint & lunar constraint \\
\colrule
\\
$g_{\rm P}^e g_{\rm S}^N/(\hbar c)$  & $(-3.5\pm 8.5)\times 10^{-37}$ & $(+0.2\pm 1.6)\times 10^{-34}$ \\
${f_\perp}/(\hbar c)$                & $(-0.1\pm 2.1)\times 10^{-32}$ & $(-1.1\pm 8.6)\times 10^{-29}$ \\
$g_{\rm A}^e g_{\rm V}^N/(\hbar c)$                      & $(+0.2\pm 1.2)\times 10^{-56}$ & $(-3.1\pm 2.4)\times 10^{-50}$ 
\end{tabular}
\end{ruledtabular}
\label{tab: Dobr params}
\end{table}
%
%
%	Figure 14
%
%
%
%	Figure 15
%

\subsection{A test of the Equivalence Principle for spin?}
It may be of some interest to interpret our constraints as an ``equivalence-principle'' test for intrinsic spin, {i.e.} to ask if the gravitational mass of an electron with its spin pointing toward the sun is identical to that of an electron whose spin points away from the sun. In this case the relevant parameter is
\begin{equation}
\tilde{\eta} \equiv \frac{[m_g]_{\rm t}-[m_g]_{\rm a}}{\frac{1}{2}([m_g]_{\rm t}+[m_g]_{\rm a})}=\frac{2\beta_{\rm sun}}{V_N}~,
\end{equation} 
where the subscripts t and a refer to an electron whose spin points toward or away from the sun, $\beta_{\rm sun}$ is the component of our signal $\bm{\beta}$ that tracks the sun, and $V_N=-5.04\times 10^{-3}$ eV is the Newtonian potential energy of a laboratory electron in the field of the sun.
We find that $\beta_{\rm sun}=(-0.62 \pm 1.30) \times 10^{-22}$ eV, which leads to
\begin{equation}
\tilde{\eta}=(+2.4\pm 5.2)\times 10^{-20}~,
\end{equation}
where the errors are $1\sigma$. A considerably tighter
%, but a bit less precise,
 limit comes from considering electrons falling toward the center of our galaxy.  In this case $V_N=m_e \Theta_0^2$ where $\Theta_0=220\pm 20$ km/s is the velocity of the solar system around the center of the galaxy, which gives $V_N=-0.275\pm 0.050$ eV. In this case we find $\beta_{\rm gal}=(+0.27 \pm 1.50) \times 10^{-22}$ eV, which leads to
\begin{equation}
\tilde{\eta}=(+0.2\pm 1.1)\times 10^{-21}~.
\end{equation}
\subsection{Measurement of the spin density in SmCo$_{5}$}
The absolute ``gyrocompass'' calibration of the spin content of our pendulum in Eq.~\ref{eq: Np from gyro} provides a clean comparison of the spin density, $\rho_{\rm s}$, of room-temperature grade~22 SmCo$_5$ to that of Alnico 5,
\begin{eqnarray}
\rho_{\rm s}({\rm SmCo}_5) \!&-& \!\rho_{\rm s}({\rm Alnico})= -N_{\rm p}^{\rm exp} /(V \eta) ~~~~~~~\nonumber \\
~~~&=& (-3.66\pm 0.08)\times 10^{22}~{\rm spins/cm}^3~,~~~~~~
\label{eq: spin density}
\end{eqnarray}  
when the SmCo$_5$ and  Alnico have identical magnetizations of $B_0 = 9.6\pm 0.2$ kG. 
The quantities $\eta$ and $V$ are defined in Eq.~\ref{eq: Npol 0th approx}. We have 
neglected the very small measured magnetic moment ($\mu_B=-0.04$) of Sm in SmCo$_5$ compared to the Co moment 
($-8.97\mu_B$), so that the equality of the magnetic fields in the two materials implies that the magnetizations satisfied
$M({\rm Co})=M({\rm Alnico})$~. The spin density in Alnico is
\begin{eqnarray}
\rho_{\rm s}({\rm Alnico})&=& B_0/(\mu_0 \mu_B))f_{\rm Alnico} \nonumber \\
&=& (7.85 \pm 0.17)\times 10^{22}~{\rm spins/cm}^3~,
\end{eqnarray}
where $f_{\rm Alnico}$ is taken from Eq.~\ref{eq: f Alnico}. Therefore
\begin{equation}
\rho_{\rm s}({\rm SmCo}_5) = (4.19 \pm 0.19)\times 10^{22}~{\rm spins/cm}^3~.
\end{equation}
\section{Summary}
We have shown that a torsion balance fitted with a spin pendulum can achieve
a constraint of $\sim 10^{-22}$ eV on the energy required to flip an electron spin about an arbitrary direction fixed in inertial space. This is comparable to the electrostatic energy of two electrons separated by 90 AU. We then use this and related
constraints to set sensitive limits on preferred-frame effects and non-commutative spacetime geometries. 
Our upper limits on the electron's rotation-noninvariant parameters $|\bm A^e_{X,Y}| \leq 1.5 \times 10^{-22}$~eV and  $|\bm A^e_{Z}| \leq 44 \times 10^{-22}$~eV are substantially smaller than previous work\cite{ho:03} and up to 5 orders of magnitude below the 
benchmark value $m_e^2/M_{\rm Planck}=2\times 10^{-17}$ eV. 
Corresponding constraints on preferred-frame effects involving protons and neutrons are given in Refs.~\cite{ph:01} and \cite{ca:04}. 
Interpreting our results as a constraint on non-commutative geometries we find that the minimum ``observable'' area is $|\Theta_{XZ}|$ or $|\Theta_{Y,Z}| \leq 4.9 \times 10^{-59}$ m$^2$, which corresponds to a length scale $\ell=355\;l_{\rm GUT}$ where $l_{\rm GUT}=\hbar c/(10^{16}~{\rm GeV})$. (Our limit $|\Theta_{XY}| \leq 1.5\times 10^{57}$ m$^2$ is weaker because it is derived from a lab-fixed signal which has a larger scale factor uncertainty.) These limits assume that the electron remains point-like up to an energy scale of 1 TeV.

By analyzing 17 months of data, we obtained the first constraint
on all 9 components of the Lorentz violating ``boost''  parameter $\bm{C}$ for any particle. Our upper limits on the absolute values of these components ranges between $2\times 10^{-18}$ and $23 \times 10^{-18}$~eV. To facilitate SME analyses, we also present constraints on $\bm{C}^{\prime}$,
the tensor helicity with respect to the sun. Our electron constraints have roughly the same sensitivity as related neutron constraints obtained with a dual-gas maser\cite{ca:04}.

Our constraints on $CP$-violating monopole-dipole interactions, derived from laboratory-fixed and solar-source signals, improve on previous work by factors of up to $10^4$, while our results for interactions between the spin pendulum and the sun or moon provide the first sensitive test for exotic spin-and-velocity-dependent interactions of electrons.

Finally, our observation of the gyrocompass effect on the electron spins provides a precise result bearing on the electronic spin structure of Sm$\:$Co$_5$. If we assume that effects from exotic physics are negligible, our measurements determine the density of polarized electrons in 
Sm$\:$Co$_5$ is $(4.19\pm 0.19)\times 10^{-22}\;{\rm cm}^{-3}$ at a field of 9.6 kG. 

We are now taking a new series of measurements that probe the interaction between two spins. This work is motivated by the recent development of a consistent model of gravity with spontaneous Lorentz-symmetry violation\cite{ar:04} that predicts dramatic new spin-spin forces. These new
experiments also probe the velocity-independent spin-spin forces\cite{do:06} that can arise from generic boson exchange.

\acknowledgments
Michael Harris and Stefan Bae\ss ler  developed earlier versions of this apparatus. Their work provided us with essential experience that made this result possible. 
Jens Gundlach and CD Hoyle made useful comments about this experiment and Erik Swanson helped prepare some of the figures. We thank Nilendra Deshpande, Bogdan Dobrescu, David Kaplan,
Alan Kosteleck\'y and Jesse Thaler for inspiring conversations, and Tom Murphy, Jr for advice on the astronomical calculations. This work was 
supported by NSF Grants PHY0355012 and PHY0653863 and by DOE funding for the Center for Experimental Nuclear Physics and Astrophysics. CEC is grateful for an NSF Fellowship.
\appendix
\section{Spin contributions to the magnetization in S\lowercase{m}$\,$C\lowercase{o}$_5$ and A\lowercase{lnico} 5}
\label{ap: spin content}
\subsection{Alnico 5}
The  composition of Alnico 5 by weight is 51\% Fe, 24\% Co, 14\% Ni, 8\% Al and 3\% Cu 
and its magnetization arises primarily from the polarized spins of 3d electrons in the Fe and Co. The orbital moment of these electrons is small because of quenching in the inhomogeneous crystalline electric fields\cite{bo:51}. The polarized spins are confined to small, needle-shaped regions of Fe-Co alloy that precipitate out during heat treatment. Although we are not aware of direct measurements of the spin and orbital moments of Alnico 5,  
we can infer the spin fraction from  ``Einstein-De Haas'' measurements\cite{sc:69} of the magneto-mechanical factors, $g^{\prime}$, of magnetically ``softer'' Fe-Co alloys. The $g^{\prime}$ measurements determine, $f$, the fractional spin contribution to the magnetic moment via
\begin{equation}
f=\frac{2(g^{\prime}-1)}{g^{\prime}}~.
\end{equation}
Table \ref{tab: spin fraction} gives the measured spin fractions in various Fe-Co alloys. We adopt the mean and standard deviation of these values for the Alnico spin fraction,
\begin{equation}
f_{\rm Alnico}= 0.953 \pm 0.005
\label{eq: f Alnico}
\end{equation}
%
%	Table XI
%
\begin{table}
\caption{Magneto-mechanical factors of Fe-Co alloys from Ref.~\cite{sc:69}. Percentages of Fe and Co are by weight.}
\begin{ruledtabular}
\begin{tabular}{cccc}
\% Fe  &  \% Co    &  $g^{\prime}$ & $f$ \\
\colrule
 \\
75  &   25   &  $1.918 \pm 0.002$  & 0.957 \\
50  &   50   &  $1.916 \pm 0.002$  & 0.956 \\
25  &   75   &  $1.902 \pm 0.002$  & 0.949 \\
\end{tabular}
\end{ruledtabular}
\label{tab: spin fraction}
\end{table}
\subsection{Sm$\,$Co$_5$}
\subsubsection{Co in Sm$\,$Co$_5$}
Although the magnetization of cobalt in Sm$\,$Co$_5$, as that of Alnico, arises primarily from polarized 3d electrons, the orbital moments are not as fully quenched because the rare earth atom alters the crystal lattice structure.   
Neutron\cite{gi:79,gi:83,sc:80} and photon\cite{ko:97} scattering studies, as well as NMR experiments\cite{yo:88}, have shown that the Co atoms in R$\,$Co$_5$ (R is one of the rare earth elements) occupy two different crystal sites; 2 atoms in the 2c site (Co$_{\rm I}$) and 3 in the 3g site (Co$_{\rm II}$). Although the spin fraction of the Co magnetization in  Sm$\,$Co$_5$ has not been measured, experiments consistently find that the Co magnetic moments in  R$\,$Co$_5$ compounds are independent of temperature and, as shown in Table~\ref{tab: spin fraction Co}, vary only slightly 
across the rare earth series. We assume that the Co spin fractions also vary only slightly with R to infer 
$f({\rm Co_I})$ and $f({\rm Co_{II}})$ for Sm$\;$Co$_5$ as given in Table~\ref{tab: spin fraction Co}.
\begin{table}
\caption{Magnetic moments and spin fractions for RCo$_5$ compounds. Unless otherwise noted results are deduced from neutron scattering measurements\cite{sc:80,gi:83}.}
\begin{ruledtabular}
\begin{tabular}{lcccc}
compound  & $\mu({\rm Co_I}$)  &  $f({\rm Co_I})$ &   $\mu({\rm Co_{II}})$ & $f({\rm Co_{II}})$ \\
\colrule
 \\
Y$\,$Co$_5$\footnotemark[1]   &  $-1.77(2)\;\mu_{\rm B}$   &  0.74(5)  & $-1.72(2)\;\mu_{\rm B}$ & 0.84(4) \\
Ce$\,$Co$_5$\footnotemark[2]  &                            &  0.76     &                         & 0.90   \\
Sm$\,$Co$_5$\footnotemark[3]  &  $-1.86(2)\;\mu_{\rm B}$   &  0.72(6)\footnotemark[4]  & $-1.75(2)\;\mu_{\rm B}$ &  0.85(6)\footnotemark[4]       \\
Nd$\,$Co$_5$\footnotemark[3]  &  $-1.95(3)\;\mu_{\rm B}$   &  0.67(5)  & $-1.90(3)\;\mu_{\rm B}$ & 0.80(3)  
\end{tabular}
\end{ruledtabular}
\footnotetext[1]{Ref.~\cite{gi:83,sc:80}}
\footnotetext[2]{NMR measurement from Ref.~\cite{yo:88}}
\footnotetext[3]{Ref.~\cite{gi:83}}
\footnotetext[4]{value inferred from neighboring compounds}
\label{tab: spin fraction Co}
\end{table}
The total spin moment of the 5 Co atoms 
\begin{eqnarray}
\mu_{\rm S}({\rm Co})&=&2f({\rm Co_I})\;\mu({\rm Co_I})+3f({\rm Co_{II}})\;\mu({\rm Co_{II}}) \nonumber \\ 
&=&(-7.14\pm 0.39)\,\mu_{\rm B}~, \nonumber \\
\mu({\rm Co})&=&2\mu({\rm Co_I})+3\mu({\rm Co_{II}}) \nonumber \\ 
&=&(-8.97\pm 0.10)\,\mu_{\rm B}~, 
\label{eq: mu_s Co}
\end{eqnarray}
which implies that 
\begin{equation}
f_{\rm Co} = \mu_{\rm S}({\rm Co})/\mu({\rm Co})=0.80\pm0.04~.
\label{eq: f Co}
\end{equation}

\subsubsection{Sm in Sm$\,$Co$_5$}
The Sm atoms in Sm$\,$Co$_5$ are in the $3^+$ ionic state with a $(4f)^5(6s)^2$ electronic configuration. Hund's rules predict that the free Sm$^{3+}$ ion's orbital momentum has the largest possible antisymmetric value of $L=5$ and that the spins are in the symmetric $S=5/2$ state. Therefore one expects large orbital as well as spin contributions to the magnetic moment. One also expects that the ion's ground state has $S$ and $L$ coupled to the minimum $J=5/2$. This simple model predicts $\mu=-0.714\;\mu_{\rm B}$, $\mu_S=+3.57\;\mu_{\rm B}$ and $\mu_L=-4.29\;\mu_{\rm B}$. However,
the Sm ions are embedded in a lattice and the crystalline and exchange fields alter this simple picture. Neutron scattering experiments\cite{gi:79} on Sm$\,$Co$_5$ find that $\mu({\rm Sm})=-0.38\,\mu_{\rm B}$ at a temperature of 4.2K. However, at room temperature, $k_{\rm B}T$ becomes comparable to the splitting of the magnetic substates
and thermal population of the various $M_J$ states causes the measured
Sm magnetic moment to be very small, $\mu({\rm Sm})=-0.04\,\mu_B$\cite{gi:79}.
Using the Sm wavefunctions in Ref.~\cite{gi:79}, we calculate that at the temperature maintained during our experiments
\begin{equation}
\mu_{\rm S}({\rm Sm})=+3.56\mu_B
\label{eq: mu_s Sm}
\end{equation}

The X-ray magnetic Compton scattering measurement reported in Ref.~\cite{ko:97} probes the spin moment of Sm in Sm$\,$Co$_5$ directly, but provides no information about the total magnetic moment.  The ratio of Co to Sm spin moments from this measurement, $R=-0.23\pm0.04$, together with Eq.~\ref{eq: mu_s Co}, corresponds to
\begin{equation} 
\mu_{\rm S}({\rm Sm})=+1.67\pm0.31\;\mu_B~.
\end{equation}
\subsubsection{Ratio of Sm to Co spin moments}
Our calculation of the number of polarized electrons in our pendulum, Eq.~\ref{eq: Npol}, depends on $R$, the ratio of Sm to Co spin moments. The neutron scattering results in Eqs.~\ref{eq: mu_s Sm} and \ref{eq: mu_s Co} give
\begin{equation}
R\equiv \mu_{\rm S}{\rm (Sm)}/\mu_{\rm S}{\rm (Co)}=-0.50~.
\end{equation}
On the other hand the photon scattering work\cite{ko:97} directly gives
\begin{equation}
R=-0.23\pm 0.04~.
\end{equation}
We accomodate this difference by assuming that $R$ is equally likely to lie anywhere between these two values.
\section{``Feetback'' tilt elimination system}
\label{ap: feetback}
\begin{figure}[ht]
\hfil\scalebox{.49}{\includegraphics*[56pt,26pt][483pt,699pt]{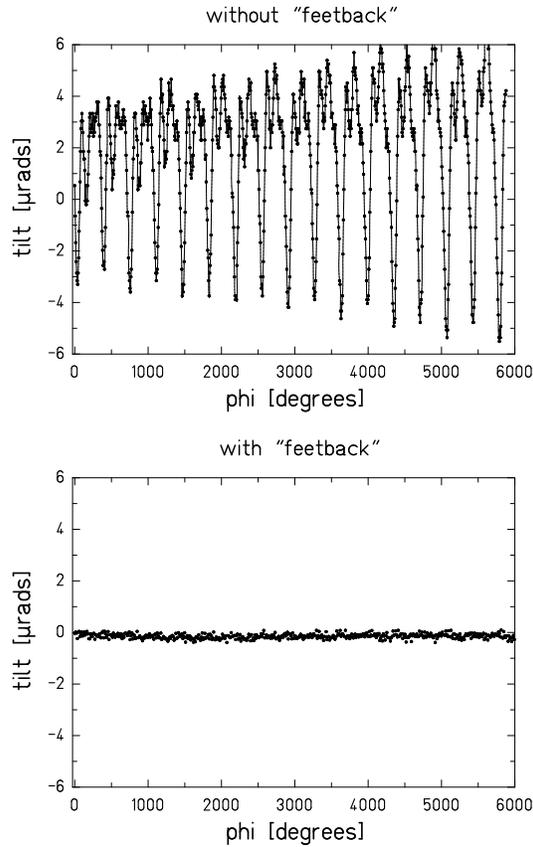}}\hfil
\caption{Performance of the ``feetback system'' showing the tilt measured by co-rotating level sensors. Upper panel: ``feetback'' off. The rapid
fluctuations arise from imperfections in the turntable bearing, slower variations arise from the varying tilt of the laboratory floor. Lower panel: ``feetback'' on.}
\label{fig: feetback}
\end{figure}

We developed a
digital feedback system that continuously corrects for the varying tilt of the laboratory floor and imperfections in the turntable itself. It analyses inputs from an orthogonal pair of co-rotating electronic tilt sensors and, every 2 complete revolutions of the turntable, computes the tilt as a function of turntable angle and cancels this predicted tilt by  
controlling the length of feet that support the turntable.
The performance of the leveling system is shown in Fig.~\ref{fig: feetback} which demonstrates that the system compensates not only for the slowly-varying tilt of the laboratory floor but also for the more rapid variations caused by imperfections in the turntable bearing.

%
%	Figure 16
%

\begin{figure}[h]
\hfil\scalebox{0.85}{\includegraphics*[214pt,57pt][390pt,231pt]{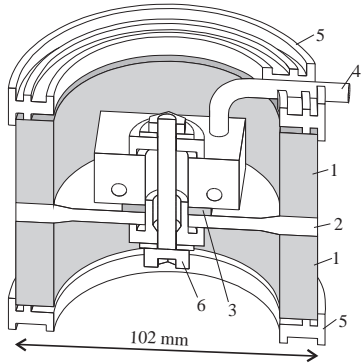}}\hfil
\caption{Cross section of a foot of the E\"{o}t-Wash rotating torsion balance.}
\label{fig: feet design}  

\end{figure}

The tilt is measured by %two perpendicular 
Applied Geomechanics Inc. inclinonometers (AGI's) 
mounted on the rotating apparatus close to the upper attachment point of the pendulum's suspension fiber.
A constant tilt of the rotation axis appears as a 1$\omega$ signal on each AGI, with a $90^{\circ}$ phase difference 
between the two units. The analog signals of the AGI's are digitized by the data acquisition system and analyzed to determine the tilt with a precision of $\approx 1~\mu{\rm rad}
/\sqrt{\rm Hz}$. This tilt is then removed by adjusting the lengths of 2 of the 3 feet upon which the apparatus rests.
The lengths of the two feet are 
are controlled 
by varying their temperatures.
Figure~\ref{fig: feet design} shows a cross section through one foot. The expanding and shrinking components of a foot consist of two lead rings (1) which are soldered to a copper disk (2). Thermal energy can be pumped into or out of the copper disk by a Peltier element (3) thermally coupled to a brass block (4) that is 
%. The brass block is 
held at constant temperature by circulating water from a temperature-stabilized reservoir. Two  G10 rings (5) thermally isolate the lead rings from the laboratory floor and from a stainless steel disk on top, upon which the apparatus rests. The Peltier element and the brass block are clamped to the copper disk by one bolt (6). Specially formed G10 pieces provide thermal isolation between the bolt, the copper disk and the brass block.
%
%	Fig 17
%

\begin{widetext}
\begin{figure}[h!]
\hfil\scalebox{0.85}{\includegraphics*[63pt,5pt][543pt,277pt]{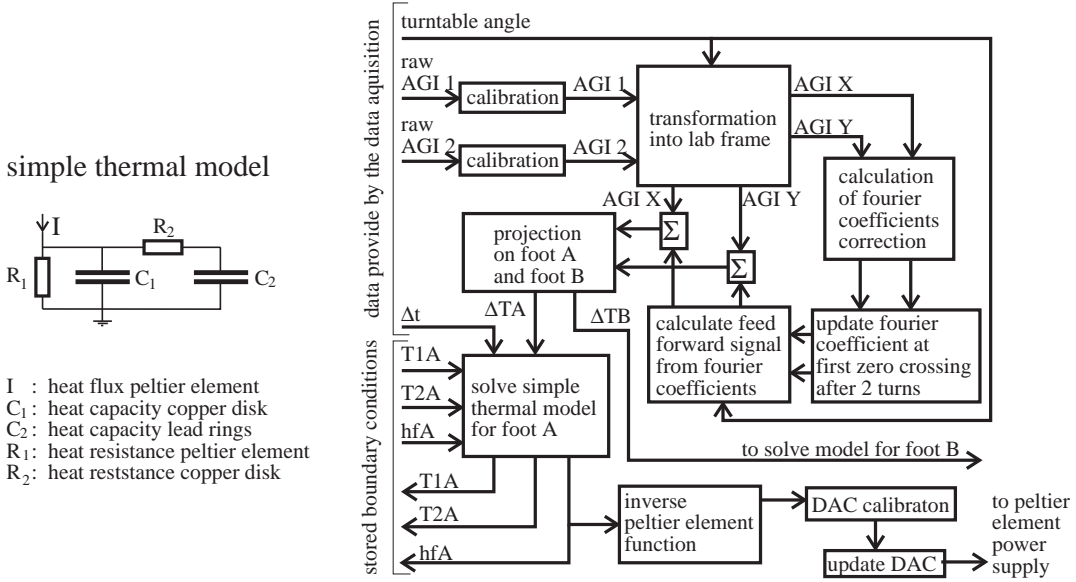}}\hfil 
\caption{Thermal model and flow chart of the simplified feetback loop implemented at the E\"{o}t-Wash II rotating torsion balance}
\label{fig: thermal model} 
\end{figure} 
\end{widetext}

The thermal capacity and thermal resistance of the copper disk and the lead rings produce a 
14-second delay in the expansion of the lead rings compared to the change in the heat flux provided by the Peltier element. In addition, the 
thermal conductivity of the Peltier element, which connects the copper disk to the temperature reservoir, affects the time 
constant of the response. Taking these factors into account, we developed a model, shown in Fig.~\ref{fig: thermal model}, to predict
a foot's response to a change in heat flux delivered by its Peltier element.
The model poses no problem for a real-time calculation because it can be solved analytically. The delayed response of the feet together with an 8-second integration time of the AGI's leads to a low-pass behavior of the leveling system. Therefore the response of the feet to fast changes of the tilt caused by imperfections of the ball bearings in the 
turntable is limited. The tilt caused by the imperfections of the bearing is periodic in the turntable angle. This allowed us to express the bearing tilt in Fourier coefficients of the turntable angle. Once these Fourier coefficients are calculated, they are fed forward to compensate for the bearing imperfections. A simplified flow chart of the feedback loop is shown in Fig.~\ref{fig: thermal model}.
With the feetback switched on, a typical tilt of a few $\mu$rad per day is reduced to less than $10~$nrad.

After the simple model described above was implemented, we found it necessary to add one further refinement. The DC offsets
of the AGI's drift slowly in time. A feedback system that holds constant the AGI signals does not guarantee true level.
With the feedback on, a drifting offset produces a 1$\omega$ AGI signal with the opposite phase
lag between the two sensors compared to a true tilt. Our model extracts this Fourier component and uses it to
determine the AGI readings that correspond to true level. 

%
%	Fig 18
%
\newpage

\end{document}